\documentclass[modern]{aastex631}
\usepackage{natbib}
\usepackage{appendix}

\usepackage{comment}
\usepackage[english]{babel}
\usepackage[autostyle]{csquotes}

\usepackage{url,hyperref,microtype}

\newcommand{\unit}[1]{\,\mathrm{#1}}

\begin{document}

\title{A Detailed Analysis of a Magnetic Island Observed by WISPR on \textit{Parker Solar Probe}}

\author[0000-0001-7220-6583]{Madison L. Ascione}
\affiliation{George Mason University 4400 University Dr. Fairfax, VA 22030, USA}

\author[0000-0002-6420-8404]{Angel J. Gutarra-Leon}
\affiliation{George Mason University 4400 University Dr. Fairfax, VA 22030, USA}

\author[0000-0002-3089-3431]{Shaheda Begum Shaik}
\affiliation{George Mason University 4400 University Dr. Fairfax, VA 22030, USA}

\author[0000-0002-4459-7510]{Mark Linton}
\affiliation{US Naval Research Laboratory 4555 Overlook Avenue SW, Washington, DC 20375, USA}

\author[0000-0002-8692-6925]{Karl Battams}
\affiliation{US Naval Research Laboratory 4555 Overlook Avenue SW, Washington, DC 20375, USA}

\author[0000-0002-5068-4637]{Paulett C. Liewer}
\affiliation{Jet Propulsion Laboratory 4800 Oak Grove Dr, Pasadena, CA 91109, USA}

\author[0000-0002-8353-5865]{Brendan Gallagher}
\affiliation{US Naval Research Laboratory 4555 Overlook Avenue SW, Washington, DC 20375, USA}

\begin{abstract}
We present the identification and physical analysis of a possible magnetic island feature seen in white-light images observed by the Wide-field Imager for Solar Probe (WISPR) on board the \textit{Parker Solar Probe} (\textit{Parker}). The island is imaged by WISPR during \textit{Parker’s} second solar encounter on 2019 April 06, when \textit{Parker} was $\sim$38 R$_{\odot}$ from the Sun center. We report that the average velocity and acceleration of the feature are approximately $334 \unit{km\,s^{-1}}$ and $-0.64 \unit{m\,s^{-2}}$. The kinematics of the island feature, coupled with its direction of propagation, indicate that the island is likely entrained in the slow solar wind. The island is elliptical in shape with a density deficit in its center, suggesting the presence of a magnetic guide field. We argue that this feature is consistent with the formation of this island via reconnection in the current sheet of the streamer. The feature’s aspect ratio (calculated as the ratio of its minor to major axis) evolves from an elliptical to a more circular shape that approximately doubles during its propagation through WISPR's field of view. The island is not distinct in other white-light observations from the \textit{Solar and Heliospheric Observatory} (SOHO) and the \textit{Solar Terrestrial Relations Observatory} (STEREO) coronagraphs, suggesting that this is a comparatively faint heliospheric feature and that viewing perspective and WISPR's enhanced sensitivity are key to observing the magnetic island.
\end{abstract}


\section{Introduction} \label{sec:Introduction}
The \textit{Parker Solar Probe} launched in August 2018 on a seven-year, 24-perihelion mission to provide progressively closer encounters with the Sun, with \textit{Parker’s} first perihelion at $35$ solar radii (R$_{\odot}$) and final perihelion less than ten solar radii from the Sun center. The Wide-field Imager for Solar Probe \cite[WISPR;][]{2016Vourlidas} is the only imaging instrument onboard the probe and is equipped with two cameras. The cameras observe at a fixed angular field that extends a total of $95^{\circ}$ radially in elongation and $50^{\circ}$ in the transverse direction; WISPR Inner (WISPR-I) approximately covers $13.5^{\circ}$ – $53.0^{\circ}$ in elongation and WISPR Outer (WISPR-O) approximately covers $50.5^{\circ}$ – $108.5^{\circ}$. The cameras provide broadband, white-light heliospheric images designed to observe coronal structures and outflows over an evolving field of view. WISPR’s sensitivity and proximity to the Sun provide new understanding to previously studied solar activity and insight into fine structures within the visible light corona, which consists of two components: light scattered by free electrons, commonly called the K-corona, and light scattered by interplanetary dust, commonly called the F-corona \citep{1998Kimura}.

WISPR is designed to observe visible light structures and solar outflow and is currently partway through its $19^{th}$ orbit. Observations have provided insight into the internal structure of a slow, streamer blowout coronal mass ejection (CME) \citep{2020Hess} and a closer look at finer substructures inside streamer rays that display a strong connection to the origin of the heliospheric plasma sheet \citep{2020Poirier, 2023Liewer}. WISPR has also proven capable of identifying faint inner solar system dust structures, such as the circumstellar dust ring in Venus’ orbit \citep{2021Stenborg_Venus} and the dust-depletion zone near the Sun \citep{2021Stenborg_depletion}. Additionally, WISPR has captured the first white-light detection of the dust trail following the orbit of asteroid 3200 Phaethon \citep{2020Battams, 2022Battams}. These new findings emphasize the high sensitivity provided by the WISPR cameras and the advantageous proximity of \textit{Parker} to the Sun. The sensitivity and proximity of WISPR have led to another exciting observation during \textit{Parker Solar Probe's} second perihelion; as noted in \cite{2019Howard}, WISPR observed an oblong structure within a streamer, consistent with that of a two-dimensional (2-D) magnetic island. This magnetic island feature is the focus of this study, as WISPR’s observation of a potential magnetic island feature indicates newly detected dynamics within the coronal streamer. 

Many previous studies have observed outward-moving density enhancements, otherwise known as streamer “blobs”, that have been hypothesized to be magnetic islands \citep{1999Einaudi, 2003Ko, 2024Cappello}. Blobs have been observed in white light coronagraphs with the Large Angle Spectrometric Coronagraph \cite[LASCO;][]{1995Brueckner_LASCO} on the \textit{Solar and Heliospheric Observatory} \cite[SOHO;][]{1995Domingo_SOHO} and by both white light coronagraphs and heliospheric imagers on the Sun-Earth Connection Coronal Heliospheric Investigation \cite[SECCHI;][]{2008Howard_SECCHI} on the \textit{Solar Terrestrial Relations Observatory} mission \cite[STEREO;][]{2008Kaiser}. Blobs have been observed to move radially outward from the tips of helmet streamers, doubling their speed from $150 \unit{km\,s^{-1}}$ around $5$ R$_{\odot}$ to $300 \unit{km\,s^{-1}}$ near $25$ R$_{\odot}$ \citep{1997Sheeley, 2009Sheeley}, with acceleration typically between $3.4-5.5 \unit{m\,s^{-2}}$ \citep{1998Wang}. A subclass of streamer blobs observed by LASCO and SECCHI instruments appear to have inward components towards the Sun, also often referred to as “in-out pairs” or blobs with raining inflows \citep{2007Sheeley&Wang, 2017Sanches-Diaz}.

Magnetic islands have been predicted to form via tearing-mode instability by reconnections of oppositely directed magnetic fields in a current sheet \citep{1963Furth}. Analysis of in situ \textit{Parker} data has shown evidence of reconnection at the heliospheric current sheet \citep{2020Lavraud, 2021Phan}. In 2-D, magnetic islands are proposed to be a collection of roughly elliptical magnetic field lines that retract towards a more circular shape, enclosing regions of density that continue to collect as more field lines are pulled into the current sheet and reconnect \citep{2001Einaudi, 2005Rappazzo}. The collection of field lines creates an elliptically shaped magnetic feature - the magnetic island - within the current sheet. The two-and-a-half dimensional (2.5-D) simulations conducted by \cite{2005Rappazzo} explore the average expansion and diamagnetic force experienced by a magnetic island when moving radially outward from the Sun along a current sheet enclosed within a helmet streamer. The simulations exemplify this formation; the island is initially elongated and thin but becomes more circular as the island evolves, and the tension in the magnetic field lines reduces. The simulations display a pronounced ring of density enhancement surrounding a core of relatively lower density. \cite{2005Rappazzo} find that the density enhancement is due to a concentration of density trapped on field lines, which annihilate. However, the core of relatively lower density is not discussed.

In 2.5-D and three dimensions (3-D), magnetic islands are thought to be helical field lines that wrap around a central guide field, also called flux ropes, with a roughly elliptical cross-section \citep{2011Daughton}. Simulations have been conducted in order to understand how the magnetic island features form within the current sheet. Figure 1a in \cite{2011Daughton} displays extended 3-D flux ropes that form in the current sheet as a result of the tearing-mode instability \citep{1963Furth}. The foreground of this figure displays four distinct plasmoid features, similar to the 2.5-D magnetic island feature discussed in \cite{2005Rappazzo}. The \enquote{islands} are, thus, a 2-D slice through the cylindrical structure of the flux rope. The island-like structure appears when the viewer is looking along the flux rope axis.

In this paper, we present and analyze the proposed magnetic island feature, noted but not analyzed in \cite{2019Howard}, as observed by WISPR during \textit{Parker Solar Probe's} second perihelion. In Section \ref{sec:WISPR Data Products and Initial Observation}, we discuss WISPR data products and the initial observation of the magnetic island in the WISPR-I images. In Section \ref{sec:Tracking Methods}, we present the two tracking methods utilized to determine the radial trajectory of the island and track the island through the WISPR images. Section \ref{sec:Analysis} presents an analysis of the orientation and dimensions of the island, as well as an analysis of the island’s orientation within the streamer and a check of the dimensions of the feature. Section \ref{sec:Results and Discussion} presents the analysis results and a discussion of the properties that distinguish the magnetic island as a subset of other previously observed streamer blobs. Finally, Section \ref{sec:Summary and Conclusion} provides a summary of the results and conclusions from this study.

\section{WISPR Data Products and Initial Observation}\label{sec:WISPR Data Products and Initial Observation}
WISPR data are released in three data products at different levels of processing; Levels 1, 2, and 3\footnote{The WISPR data is provided in the \enquote{Flexible Image Transport System,} FITS, which is a flexible file format used for storing, transporting, and analyzing astronomical data \citep{1981Wells, 2020OPossel}.}. In this paper, we utilize the Level 3 (L3) processing, which removes the smooth component of the F-corona and other instrumental artifacts and leaves emission from K-corona structures such as streamers and CMEs, as well as discrete dust features from comet dust trails and both galactic and solar objects. For a discussion of the L1 and L2 processing and an in-depth discussion of the L3 processing, see \cite{2021Hess} and the Appendix of \cite{2023Liewer}. We also utilize LW-processing, which is an alternative version of the L3 data. LW-processing uses a customized technique to remove quasi-stationary structures, such as streamers. This technique highlights discrete K-corona structures and small-scale transient features and is an additional means to inspect and analyze the data. A detailed description of the LW-processing and data products can be found in Appendix A of \cite{2022Howard}. In addition to the two processing techniques, we apply two filters to the data products: the uniform and sigma filters. The uniform filter replaces the value of a pixel with the mean value of the area centered at that pixel, and the sigma filter suppresses/removes point-like bright features such as stars and cosmic rays.

\begin{figure}
    \includegraphics[width=1\textwidth]{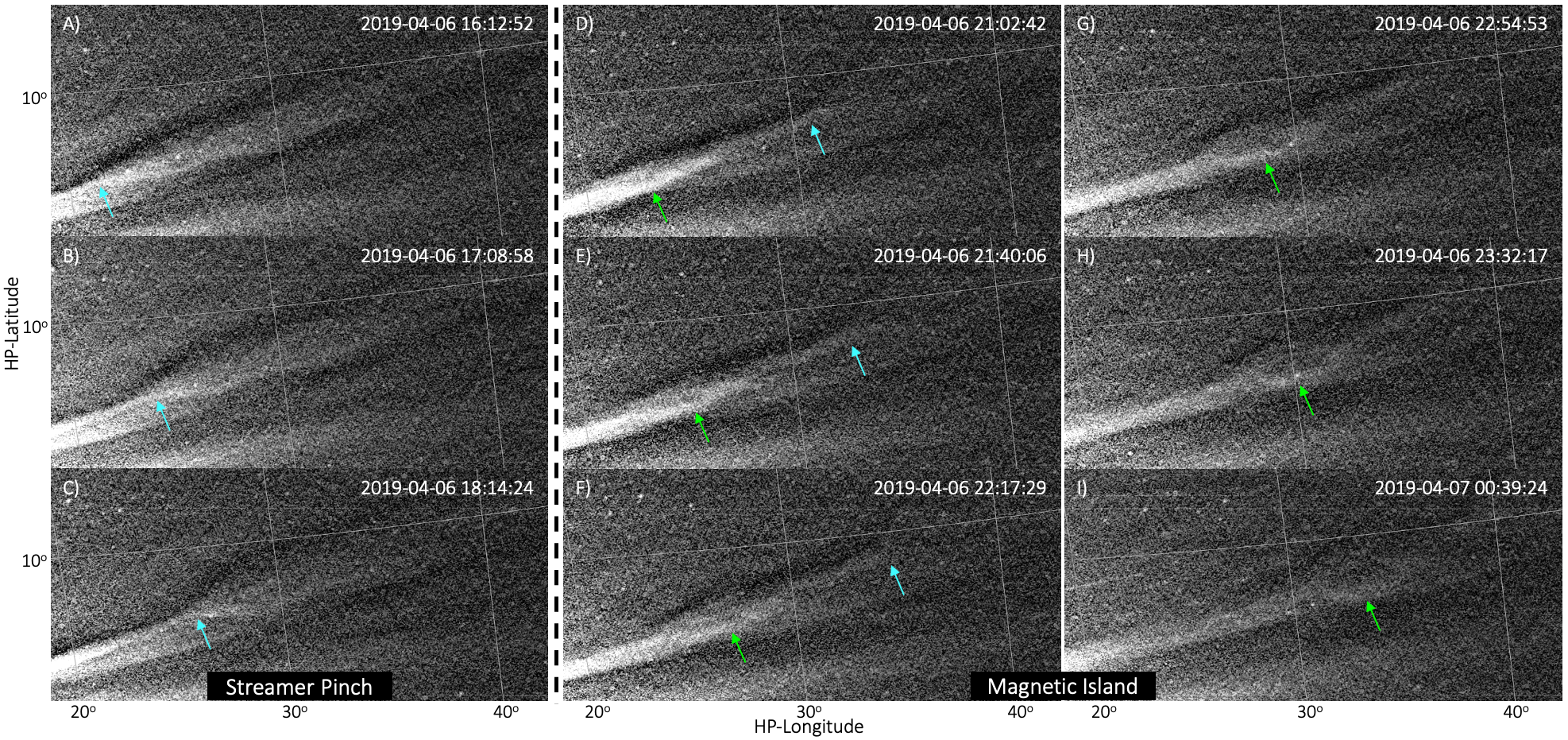}
    \caption{Initial observation of the streamer and magnetic island on 2019 April 06. Panels A-C display the change of shape in the streamer, indicated by the cyan arrow at the location of the streamer pinch. Panels D-I display the magnetic island throughout its propagation, indicated by the green arrow pointing to the lower outer edge of the island. A density deficit is observed at the center of the outer high-density ring. The cyan arrow in Panels D-F indicates the same leading streamer pinch in Panels A-C. The arrows are included as a visual aid and are placed using a by-eye approximation. The images are WISPR-I with L3-processing and an applied sigma filter.}
    \label{fig:1}
\end{figure}

Figure \ref{fig:1} displays the initial L3 observation of the magnetic island captured by WISPR-I in the Helioprojective-cartesian\footnote{Helioprojective-cartesian (HP) coordinate system is an observer-centric, 2-D coordinate system where HP-longitude measures the angle away from Sun center, increasing towards solar West, and HP-latitude measures the angle away from Sun center, increasing towards solar North.} \citep{2006Thompson} coordinate system on 2019 April 06 during \textit{Parker's} second solar encounter, when \textit{Parker} was $\sim38.5$ R$_{\odot}$ ($\sim0.18$ AU) from the Sun center. WISPR-I observed a bright streamer change shape just a few hours before observing the feature that we infer to be a magnetic island. The streamer appears to pinch, indicated by the cyan arrow in Panels A-C of Figure~\ref{fig:1}, which results in the streamer taking a fluting shape. The island follows this pinch at $\sim$21:02:42 UT and appears as a high-density elliptical ring enclosing a visible density deficit (the deficit is hereafter referred to as a cavity). As seen in Panels D-I of Figure~\ref{fig:1}, the island (indicated by the green arrow) appears to move along the coronal streamer following the initial streamer pinch (cyan arrow) and takes a more circular shape by the end of its observation period on 2019 April 07 at $\sim$00:58:34 UT. From the tracking results in Section \ref{sec:Tracking Methods}, the island approximately travels from $16.8$ to $23.5$ R$_{\odot}$ during this period. The feature is not visible in the WISPR-O L3-processed images but is visible in the WISPR-O LW-processed images. However, the island becomes too faint for quantitative analysis in the WISPR-O LW-processed images, causing the cavity to be indistinguishable from the outer density ring. For these reasons, we only investigate the WISPR-I images in this study.

\section{Tracking Methods}\label{sec:Tracking Methods}

\subsection{Determination of 3-D Trajectory of the Island using the Tracking and Fitting Technique}\label{subsec:Tracking and Fitting Method}

WISPR produces images in which the signal at each pixel results from a line-of-sight integration, and thus, there is no information on the distance of objects along the line of sight (LOS). However, based on various assumptions that can be made about observed structures, information about the location of these structures along the line of sight can be inferred. This study makes use of a Tracking and Fitting (T\&F) technique, which was developed specifically for transients observed by WISPR \citep{2020Liewer}. The T\&F technique assumes a tracked feature propagates out from the Sun at a constant speed along a line of constant longitude and latitude in a heliocentric frame of reference and makes use of the multi-viewpoint observations of coronal features resulting from \textit{Parker Solar Probe}’s rapid, highly elliptical orbit to extract the object's trajectory. This tracking method considers the motion of both the feature (here, the island) and observer (WISPR) and corrects for the inclination of the \textit{Parker} orbit plane to the solar equatorial plane, in order to determine the tracked feature’s observed motion in a sequence of images.

The position of a feature in the images is tracked manually, using a cursor to select the feature's position in a time series of images. In this case, the center of the island was tracked in a series of images spanning about 5 hours. The tracked data is then fit to analytic expressions in order to solve for the four trajectory parameters: latitude, longitude, radial velocity, and radial distance from the Sun center at the start of tracking. The analytic expressions were derived for relating the position of the transient to its projected position in a WISPR image and for how the position in the image should change as a function of time due to both the transient’s and \textit{Parker’s} changing position. Analytic expressions were derived for both the Heliocentric Inertial\footnote{The Heliocentric Inertial (HCI) system has its $z$-axis aligned with the Sun's north pole, $x$-axis aligned in the direction of the ascending node of the solar equator on the ecliptic of January 1, 2000, 12:00 Terrestrial Time in the Julian Epoch (denoted as J2000.0), and the $y$-axis completes the right-hand coordinate system. The solar ascending node is the intersection of the solar equatorial plane with the ecliptic plane.} (HCI) and the Carrington\footnote{The Carrington (CARR) system is the heliographic coordinate system that rotates in a sidereal frame every 25.38 days; the $z$-axis is aligned with the Sun’s north pole, and the $x$- and $y$-axis rotate within the sidereal period. The beginning of a new “Carrington rotation” occurs when Carrington prime meridian coincides with the central meridian as seen from Earth. The canonical zero meridian used today started after Carrington began observing the motion of sunspots. This meridian passed through the ascending node of the solar equator on the ecliptic at Greenwich mean noon on January 1, 1854 (Julian Day 239, 8220.0) \citep{1970_Almanac}.} (CARR) coordinate frames \citep{1863Carrington, 2002Franz, 2006Thompson} under the assumption of radial propagation at a constant velocity in that frame. The same tracking data set can be used to find the solution in both frames, and the (small) difference in the two solutions is used to estimate the uncertainty. Details of the technique can be found in \cite{2020Liewer}.

\begin{figure}
    \includegraphics[width=1\textwidth]{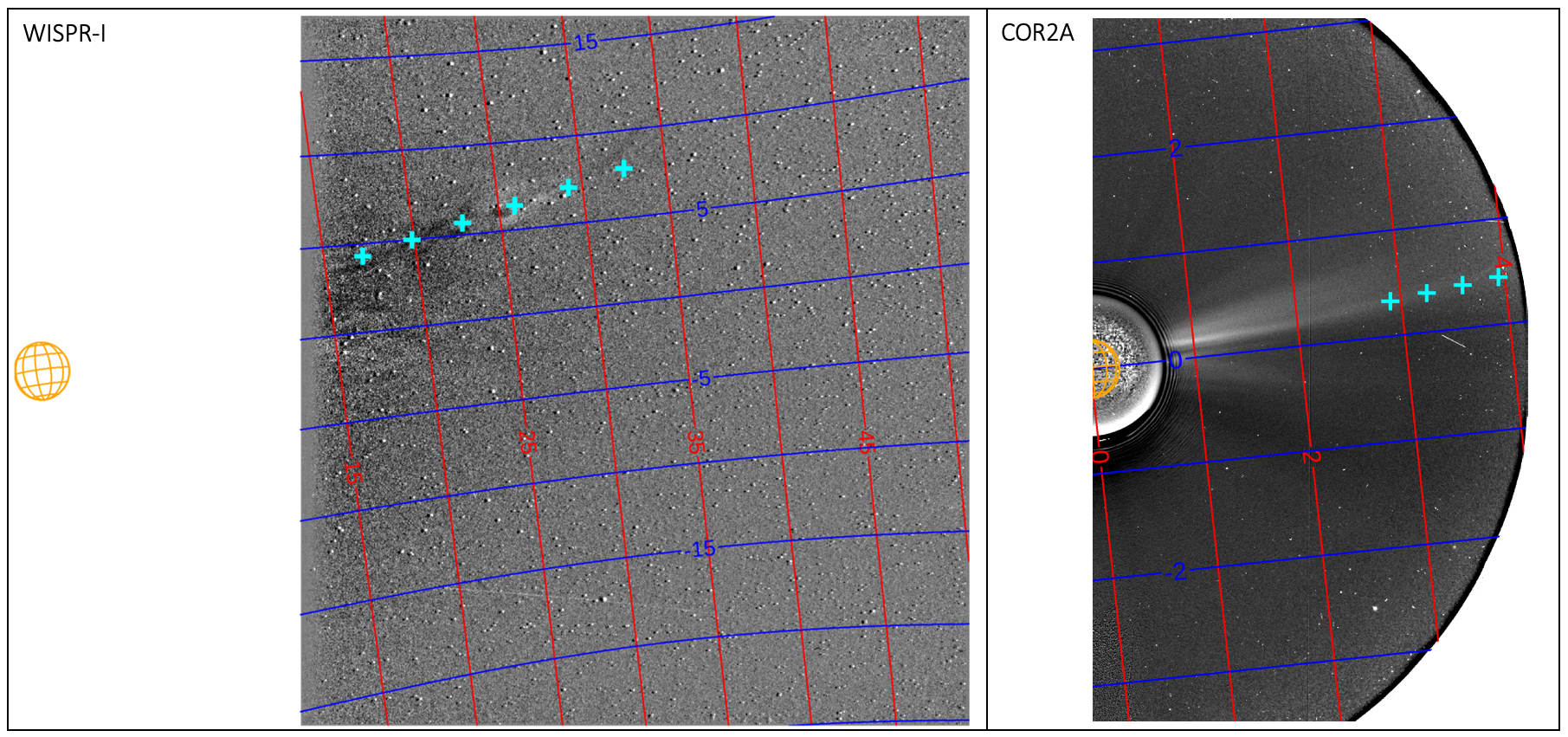}
    \caption{Projection of points along the island center trajectory onto WISPR-I and STEREO/COR2A images. The grid lines are HP longitude (red) and latitude (blue). \textit{Left:} WISPR-I LW-processed image on 2019 April 06 at 21:40 UT. The cyan points start at $r_{i}$ = $13.4$ R$_{\odot}$ at the tracking start time $t_0$ = 19:00 UT on 2019 April 06 and are calculated every hour for the next five hours using the solution velocity $v$ = $327 \unit{km\,s^{-1}}$. The fourth point is at 22:00 UT, 20 minutes after the time of the image. \textit{Right:} STEREO/COR2A image on 2019 April 06 at 06:00 UT. These are the same points as the first four in the WISPR-I image (left); the last two points are beyond the COR2A field of view. The separation of STEREO and \textit{Parker} was about $70^{\circ}$. From this viewpoint, the island appears to lie in a coronal streamer.}
    \label{fig:2}
\end{figure}

The solution determined by this technique for the trajectory of the island center was HCI longitude and latitude ($\phi, \theta$) = ($148 \pm$ $2^{\circ}$, $7 \pm$ $1^{\circ}$), $v$ = $327 \pm 4.8 \unit{km\,s^{-1}}$, and $r_{i}$ = $13.4 \pm 0.6$ R$_{\odot}$ at $t_{0}$ = 2019 April 06 at 19:00:00 UT, the time tracking started. This T\&F solution is also presented in \cite{2024Liewer} (submitted), along with the T\&F solutions and associated analyses of another three streamer blobs observed by WISPR during Encounter 2. The solution was verified by projecting 3-D locations calculated from the extracted solution back onto images used in the tracking. The left panel in Figure~\ref{fig:2} shows 3-D trajectory points (cyan) calculated hourly from the start time $t_{0}$ plotted on the LW-processed WISPR-I image on 2019 April 06 at 21:40 UT, 20 minutes before the fourth cyan point. The tracked feature is the dark island center (the cavity), a bit behind this point, and thus, the trajectory appears to match the data well. The fourth point falls exactly on the center of the island in the WISPR image 20 minutes later, but we use the earlier image so that the island center is visible in the WISPR image. Thus, the 3-D projection of the solution falls on the island feature tracked, and the solution is verified.

Because we have the 3-D solution in an inertial frame, we could also project these trajectory points onto a STEREO/COR2A \citep{2008Howard_SECCHI} image to find out how it would appear from that viewpoint \citep{2020Liewer}. COR2 is a visible light Lyot coronagraph in the SECCHI instrument suite that images the inner and outer corona from $2.5-15$ R$_{\odot}$. The island feature is expected to be along the coronal streamer as per the WISPR-I image; however, there is no visual indication of the feature in the COR2A images, as shown in the COR2A panel in Figure~\ref{fig:2}. The COR2A and WISPR images in Figure \ref{fig:2} show the same island trajectory from different viewing points. This is discussed further in Section \ref{subsec:Other Results}.

\subsection{12-Point Tracking}\label{subsec:12-Point tracking}

\begin{figure}
    \centering
    \includegraphics[width=0.99\textwidth]{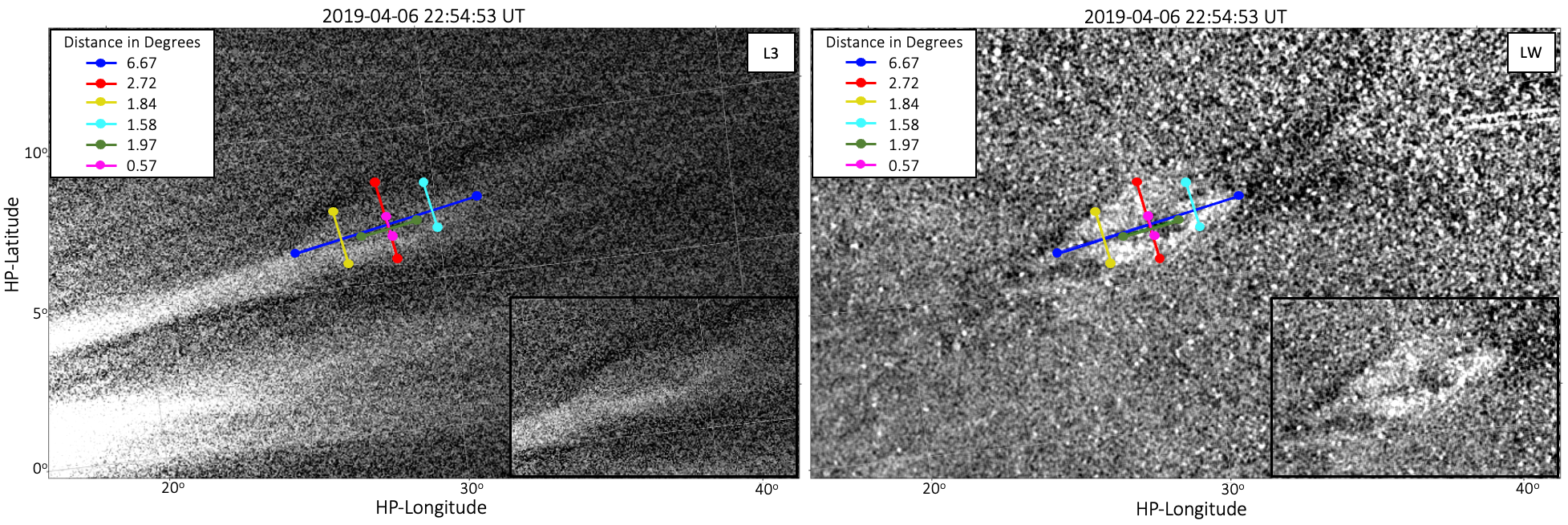}
    \caption{12-point tracking of the magnetic island at time 22:54:53 UT on 2019 April 06. The left image is L3-processed, and the right is LW-processed, both are WISPR-I images with an applied sigma filter. The two dark blue points track the leading and trailing edges of the island, and the green points track the leading and trailing edges of the cavity, respectively. The yellow, red, and cyan points were chosen as the upper and lower edges of the island feature ($1/4$, $1/2$, and $3/4$ of the way between the dark blue points), and the pink points were chosen as the upper and lower edges of the cavity ($1/2$ of the way between the green points), respectively.}
    \label{fig:3}
\end{figure}

The second way we track the island through the WISPR images is via a by-eye 12-point selection process, hereafter referred to as the 12-Point method. This method relies on human identification of the outer edges of the magnetic island feature and the outer edges of the island’s cavity, which lies within the outer density ring. Before we track the feature, we compare the shape of the island in both L3- and LW-processed images to confirm the visual presence and structure of this feature. Figure \ref{fig:3} displays a subsection of the full WISPR-I frame on 2019 April 06 at 22:54:53 UT, with both L3- (left) and LW- (right) processing. In the left L3-processed image, the island seems to protrude from the streamer with a density deficit in its center. On the right, the LW-processed image displays the island feature with the streamer subtracted. The cavity is present within the island in both the L3- and LW-processed images, although it appears much clearer in the LW-processed image. In the LW image, the presence of the island, even after removing the steamer, indicates that this feature is in and of itself independent of the main structure of the streamer. However, in the process of removing the streamer, the LW-processing may, in turn, remove part of the leading and trailing edges of the island feature. Because this processing effect may impact the overall shape of the feature, we find it necessary to reference both the L3- and LW-processed images to conduct the 12-Point tracking method.

To track the island, we identify 12 consistent points along the feature. Eight points lie on the outer edges of the island's bright density ring, and four points lie on the outer edges of the cavity, which is within the density ring (see Figure \ref{fig:3}). We determine the leading and trailing edges of both the island and cavity in the WISPR-I images and track the points. In L3-images (left panel in Figure \ref{fig:3}), the leading edge of the island is identified as the slightly rounded front tip of the density enhancement, and the trailing edge of the island appears to pinch off from the streamer (represented by the dark blue points in Figure \ref{fig:3}). Additionally, both the edges appear slightly brighter than the streamer and the inner cavity, making the edges of the island clearly identifiable. In comparison between the L3- and LW-images, the edges appear roughly in the same position even after the streamer is subtracted in the LW-image (right panel in Figure \ref{fig:3}). The cavity is dimmer than the outer density ring in both the L3- and LW-processed images. Thus, we can determine that the leading and trailing edges of the cavity lie within the edge of the bright density ring (represented by the green points in Figure \ref{fig:3}) and are roughly in line with the leading and trailing edges of the island.

To track the upper and lower edges of the island's bright density ring, we calculate the distance a quarter, half, and three-quarters of the way along the dark blue line between the leading and trailing edges. At these distances, we plot perpendicular lines to the connecting blue line to use as visual aids in order to determine the upper and lower outer edges of the island’s bright density ring. We compare the upper and lower edges between both the L3- and LW-processed images to create a full understanding of the feature. The density ring’s outer edges are represented in Figure \ref{fig:3} by the yellow, red, and cyan dots, which are, respectively, a quarter, halfway, and three-quarters of the way between the leading and trailing edges of the island. The cavity is a much smaller section of the entire feature; therefore, we only track the upper and lower outer edges of the cavity at the halfway distance between the cavity’s leading and trailing edges. Using the same process as the outer density ring, we plot a perpendicular line as a visual aid to determine the upper and lower outer edges of the cavity, which are determined to be right within the bright density ring and are represented by the pink dots in Figure \ref{fig:3}.

This 12-Point tracking method allows us to calculate the feature’s estimated physical size in R$_{\odot}$ at a specified HCI longitude. This provides insight into the dimensions of the island feature and cavity, as well as a means to estimate the velocity and acceleration. We can find the HCI Cartesian coordinates at a specified longitude of each tracked pixel. In this study, we calculate the tracked positions for the plane-of-sky and the island plane. While various plane-of-sky assumptions exist, in this study, we assume the plane-of-sky is a flat plane perpendicular to the central line of sight of the observing field of view\footnote{The center point of the FITS image is the CRPIX position, which is defined in the image header.} that intersects the Sun's center. Here, for the plane-of-sky, the WISPR-I center is at HP longitude $32^{\circ}$ and HP latitude $-4^{\circ}$. We define the island plane as the plane containing both the radial ray of travel of the island and the solar rotation axis. The HCI longitude of the radial ray of travel is provided by the T\&F tracking method. A discussion of the size and kinematic calculations is discussed in Section \ref{subsubsec:Size and Kinematics Calculation}.

\section{Analysis Methods}\label{sec:Analysis}
We utilize the results from the two tracking methods to conduct four analyses of the magnetic island feature. First, we utilize solutions from the T\&F method to understand the trajectory of the island in 3-D space, as well as determine the location of the island feature relative to \textit{Parker}, SOHO, and STEREO. Second, we utilize the points from the 12-Point tracking method to calculate the approximate physical size of the island and a rough estimation of the feature's velocity and acceleration. Third, as a check of the 12-Point method, we introduce an alternative computational process to estimate the aspect ratio and orientation of the magnetic island. This process, henceforth referred to as the Max-Point algorithm, takes advantage of the increase and decrease in intensity from the cavity to the outside of the island. Finally, we developed an algorithm to estimate the orientation of the streamer in the WISPR-I FOV. This algorithm, hereafter referred to as the Gaussian-Fit algorithm, determines the orientation of the streamer using the intensity values fitted along a series of vertical lines.

\subsection{Island Feature Orientation in 3-D Space} \label{subsubsec:Island Feature Orientation in 3-D Space}
We calculate the island’s 3-D position in space for each WISPR-I frame that observed the island between 2019 April 06 at 21:02:42 UT and 2019 April 07 at 00:58:34 UT. We use the HCI trajectory solution provided by the T\&F method to calculate the new heliocentric radius, $r_{f}$, of the island at time $t_{f}$, which is associated with one frame captured by WISPR-I:

\begin{equation}
    r_{f} = r_{i} + v*(t_{f} - t_{0}).
    \label{eq:New Radius}
\end{equation}

We use the new radius at time $t_{f}$, as well as the HCI longitude ($\phi$) and latitude ($\theta$) values of the T\&F solution, to calculate the HCI Cartesian coordinates of the feature in terms of spherical coordinates:

\begin{equation}
    x = r_{f} cos(\phi) cos(\theta),
    \label{eq:X-coord}
\end{equation}

\begin{equation}
    y = r_{f} sin(\phi) cos(\theta),
    \label{eq:Y-coord}
\end{equation}

\begin{equation}
    z = r_{f} sin(\theta).
    \label{eq:Z-coord}
\end{equation}

\begin{figure}
    \centering
    \includegraphics[width=\textwidth]{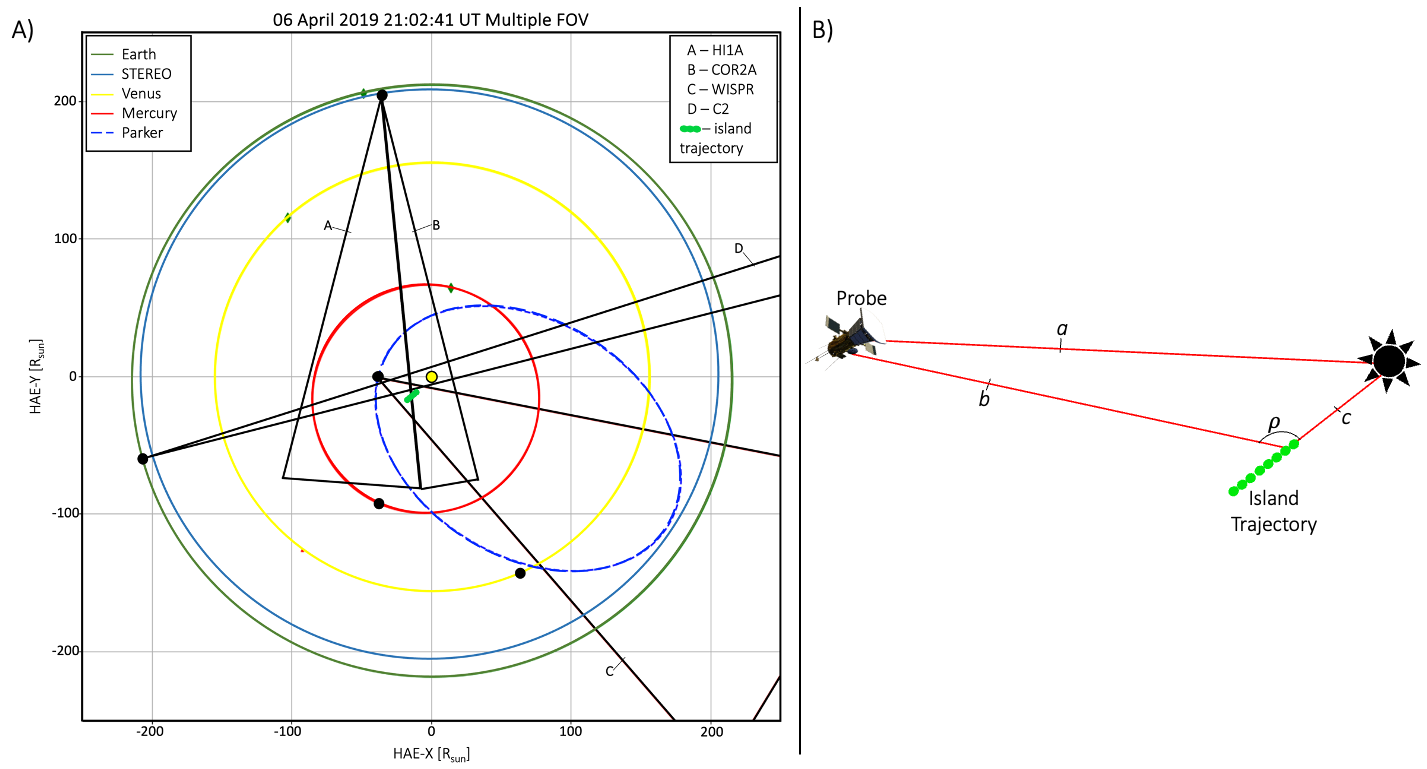}
    \caption{\textit{Left:} Position of the magnetic island, inferred from the T\&F solution, relative to the fields of view of \textit{Parker Solar Probe} (WISPR-I), SECCHI (COR2A and HI1A), and SOHO (LASCO C2) at 21:02:41 UT on 2019 April 06. The small green circles represent the radial motion of the island from 2019 April 06 at 21:02:41 UT to 2019 April 07 at 00:58:34 UT. \textit{Right:} Illustration of the viewing geometry of the magnetic island locations relative to \textit{Parker Solar Probe}. The line \textit{a} extends from the Sun center to \textit{Parker}, line \textit{b} extends from \textit{Parker} to the calculated location of island center, line \textit{c} extends from the Sun center to the calculated location of island center, and $\rho$ is the viewing angle created from the intersection between lines \textit{b} and \textit{c}.}
    \label{fig:4}
\end{figure}

We utilize the Plot Orbit tool within the PySSW software package to analyze the island's propagation relative to \textit{Parker}, SOHO, and STEREO. This analysis is conducted with the calculated Cartesian coordinates. The tool allows us to visualize the direction of propagation in both 2-D and 3-D space and helps determine whether the island feature is within the field of view of SOHO or STEREO. Figure \ref{fig:4}A shows the 2-D field of view (FOV) of WISPR-I, as well as the 2-D FOV of the COR2A coronagraph and HI1 heliospheric imager \cite[HI1A;][]{2009Eyles_HI-1} onboard STEREO-A and the LASCO C2 coronagraph \citep{1995Brueckner_LASCO} relative to the island propagation path. The plot is in the HAE-X, HAE-Y \footnote{The Heliocentric Aries Ecliptic (HAE) frame is a heliocentric coordinate system where the Z-axis normal to the ecliptic, the X-axis is oriented along the first point of Aries on the Vernal Equinox at epoch J2000.0, and the Y-axis completes the right-hand coordinate system.} plane \citep{2002Franz}, on 2019 April 06 at 21:02:41 UT, and Panel B displays a zoomed-in depiction of the island’s propagation relative to \textit{Parker}. The path of the island, represented by the line of small green dots, is roughly perpendicular to WISPR-I’s CRPIX line of sight. The dots represent the island center's position from time 21:02:42 UT on 2019 April 06 to 00:58:34 UT on 2019 April 07. In addition to the 2-D FOV plots, we check each camera's 3-D FOV plot and confirm that the island path of propagation is within the 3-D FOV of each instrument. 

\begin{figure}
    \centering
    \includegraphics[width=\textwidth]{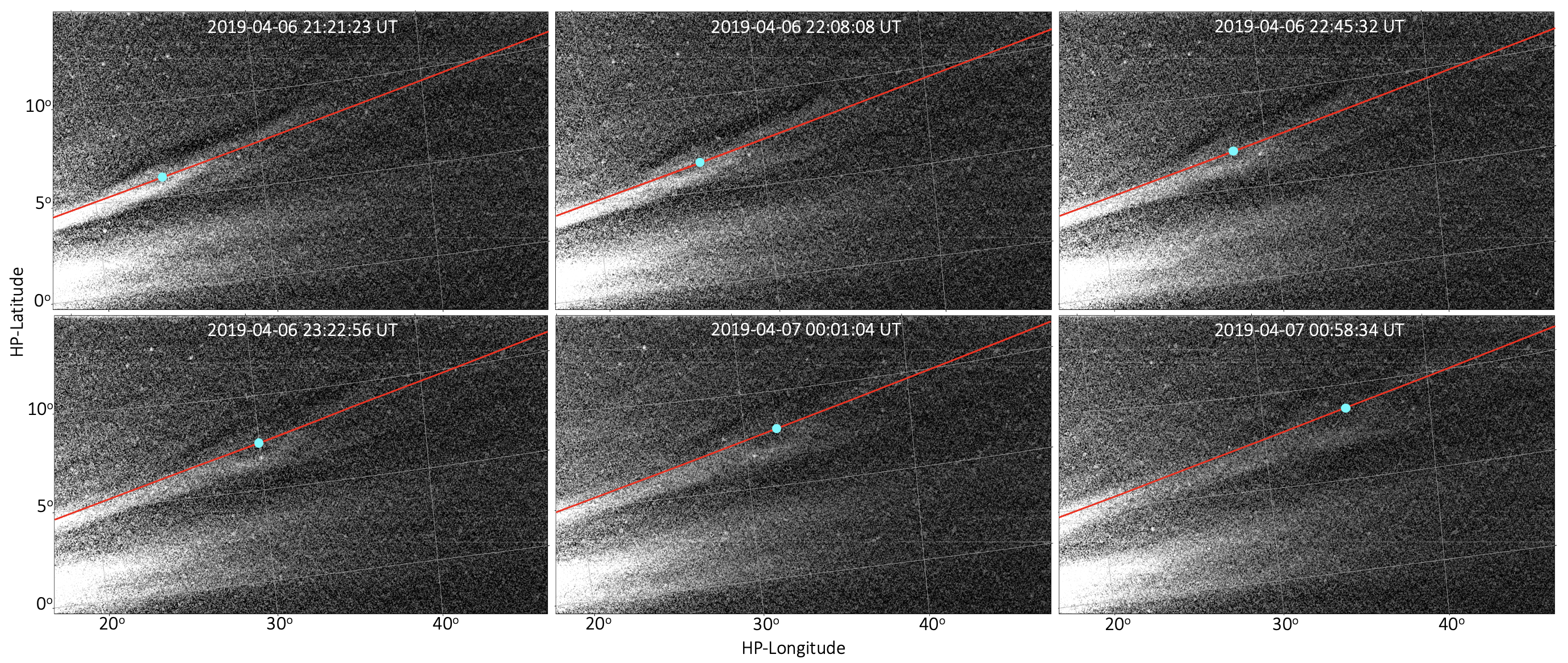}
    \caption{Overlay of calculated coordinates on L3-processed WISPR-I images. The cyan dot is the pixel location in the WISPR-I image that correlates to the 3-D island coordinate inferred from the T\&F solution. The red dashed line is the HCI longitude and latitude provided by the T\&F trajectory solution. The calculated coordinates appear to line up with the center of the magnetic island, as visually observed in the images.}
    \label{fig:5}
\end{figure}

We then plot each coordinate over the associated WISPR-I image using functions in the PySSW package to confirm that the island’s propagation within the Plot-Orbit tool lines up with the location of the island feature in the images. Figure \ref{fig:5} provides a closer look at the coordinate position relative to the island center, similar to Figure \ref{fig:2}. In Figure \ref{fig:5}, the cyan dots are the calculated 3-D coordinates at six times during the island's observation period. The red dashed line is the ray of the HCI longitude/latitude solution provided by the T\&F method. The coordinates line up well with what we understand as the center of the feature in the WISPR-I images, further verifying the tracking and fitting technique discussed in Section \ref{sec:Tracking Methods} and the results from Figure \ref{fig:2}.

Once confirming the island center coordinates, we calculate the angle at which the island is traveling relative to WISPR-I’s line of sight. This angle, which we will refer to as the viewing angle, is formed at the intersection of lines \textit{b} and \textit{c} in Figure \ref{fig:4}B. We use the law of cosines to calculate the viewing angle, $\rho$, as distances \textit{a} and \textit{c} are readily available, and \textit{b} is found by calculating the distance between the island center (obtained from the T\&F method) and \textit{Parker’s} position in space: 

\begin{equation}
    cos(\rho) = \frac{b^{2}+c^{2}-a^{2}}{2bc}.
    \label{eq:Viewing Angle}
\end{equation}

We repeat the coordinate calculation, coordinate overlay, and viewing angle calculation processes for C2, COR2A, and HI1A to gauge if these cameras observe an island-like feature or density enhancement roughly similar in shape to the magnetic island observed by WISPR-I. To do this, we determine the time range in which the island’s path lies within each FOV by increasing the overall observation period to 15:00 UT on 2019 April 06 - 04:00 UT on 2019 April 07, calculating additional coordinates, then re-uploading the 3-D coordinates to the Plot-Orbit tool. We inspect both the 2-D and 3-D FOV plots for each imager to determine the time range for which the island’s propagation lies entirely within the respective instrument’s FOV. As the time stamps are different for each coronagraph, we calculate a new set of coordinates for each time frame within the determined time range, then utilize various functions from the PySSW software package to plot the coordinates onto the corresponding images. Overall, COR2A and C2 do not observe a clear island-like feature, but HI1A does observe a density enhancement at the location of the island coordinates. The results of this process are further discussed in Section \ref{subsec:Other Results}.

\subsection{Size and Kinematics Calculation}\label{subsubsec:Size and Kinematics Calculation}
We use the points obtained through the 12-Point tracking method to calculate the approximate physical size of the island feature in the plane-of-sky and island plane, as well as analyze its kinematics along the ray of travel. In order to calculate the size or kinematics, we need to determine the HCI Cartesian position of each tracked point in the WISPR-I images. To do so, we utilize packages within the PySSW tool to first convert the tracked pixel locations to HP longitude and latitude coordinates. Then, we convert these coordinates to HCI Cartesian coordinates (in units of kilometers) at a specified longitude; $\sim 163^\circ$ HCI longitude for the plane-of-sky and $148^\circ$ HCI longitude for the island plane. 

To estimate the 2-D size of the feature, we calculate the distance between tracked points discussed in Section \ref{subsec:12-Point tracking}. We apply the standard distance equation, where a given tracked point is denoted as ($p_{x}, p_{y}, p_{z}$) in an image, and convert the distance value to solar radii\footnote{Conversion between solar radii and kilometers: 1 $R_{\odot} = 695500 \unit{km}$.}. For example, we calculate the HCI Cartesian distance between the leading and trailing edges of the island feature in an image (the dark blue line in Figure \ref{fig:3}) using the following equation:

\begin{equation}
    distance = \frac{\sqrt{\left ( p_{x_{front}} - p_{x_{rear}} \right )^{2} + \left ( p_{y_{front}} - p_{y_{rear}} \right )^{2} + \left ( p_{z_{front}} - p_{z_{rear}} \right )^{2}} }{695500}.
    \label{eq:Size Calculation}
\end{equation}

This size calculation indicates the approximate radial width and vertical thickness of the island feature's bright density ring and inner cavity. The radial width refers to the island’s 2-D major axis, signified by the dark blue line in Figure \ref{fig:3}. This represents the distance between the leading and trailing edge of the island in the radial direction in which the island travels. The vertical thickness refers to the island's 2-D minor axis, signified by the red line in Figure \ref{fig:3}, and represents the distance between the upper and lower edges of the island and is perpendicular to the radial width. These terms are used to discuss the dimensions of the island in the 2-D frame, as we do not know its extent in the third dimension.

To calculate the kinematics of the feature, we plot the height of each tracked portion of the feature from the Sun center in solar radii versus the elapsed time of the observation period. We fit a second-order polynomial to the plotted values to receive a velocity (in $\unit{km\,s^{-1}}$) and an acceleration (in $\unit{m\,s^{-2}}$). We repeat this plotting and fitting process for each of the 12 tracked points, resulting in 12 comparable velocity values, and average these twelve values for an overall velocity of the feature. Results of the size calculation and polynomial fittings are presented in Section \ref{sec:Results and Discussion}.

\subsection{Ellipse Fitting via Max-Point Algorithm}
The Max-Point algorithm is designed to find the brightest point along a radial line that originates from inside the cavity and extends past the outer edges of the island feature. This algorithm interpolates the best-fit ellipse using a direct least-squares regression method. The method places a restriction on the coefficients in order to avoid the trivial solution and ensure the best fit to the elliptical shape of the island. We applied this algorithm to LW-processed images and applied sigma and uniform filters. We cannot apply this algorithm to the L3-processed images because the intensity of the island is too faint compared to the surrounding background, and the relatively higher intensity of the streamer would interfere with the process. 

The point of origin of the radial lines influences the final shape and orientation of the ellipse; therefore, we select several starting points within the cavity and calculate a statistical average of each generated ellipse to arrive at a final ellipse. Each image is manually inspected to ensure that all the origin points are inside the cavity, and a rectangular selection box is created from within which all origin points are selected.

\begin{figure}[ht]
    \centering
    \includegraphics[width=1\textwidth]{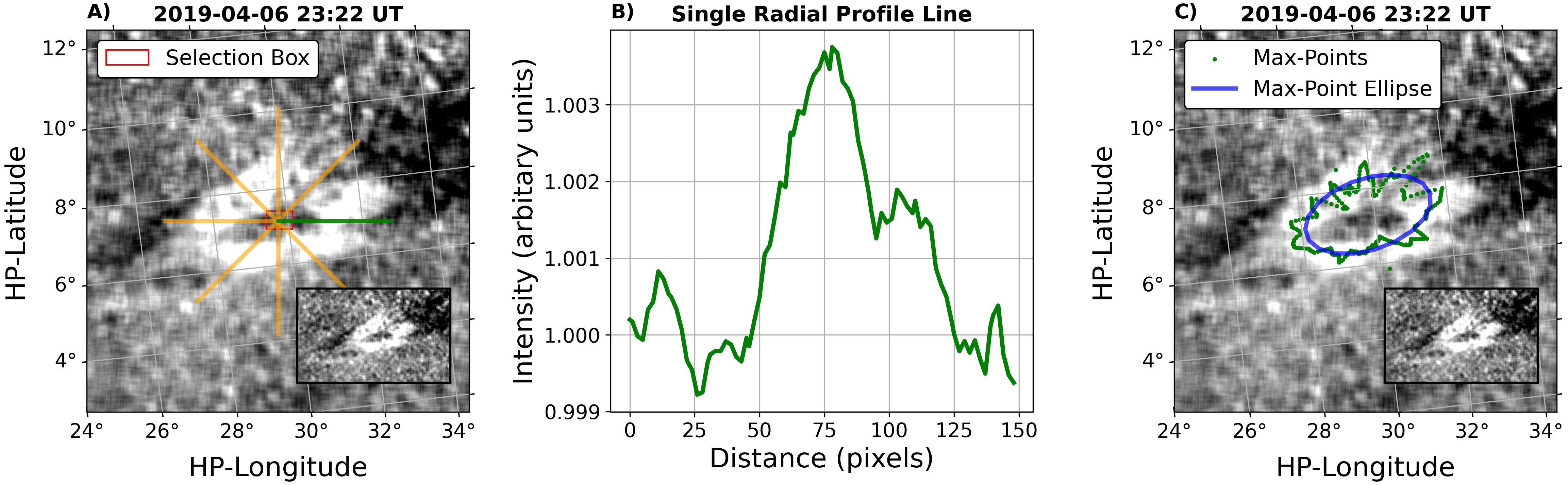}
    \caption{Steps to fit an ellipse via the Max-Point algorithm. Images shown are WISPR-I images with LW-processing. \textit{Panel A:} A pixel is chosen from the selection box (red), and radial lines (orange and green) are created to sample the bright density ring. \textit{Panel B:} The intensity profile of a single radial line (green) is displayed. \textit{Panel C:} The maximum points (green) along each radial line are plotted on the image. An ellipse (blue) is fit to the points using a direct least squares fitting algorithm.}
    \label{fig:6}
\end{figure}

Figure \ref{fig:6}A shows eight of 360 radial lines created at 1-degree increments. Each radial line originates from a single starting point within the red selection box. The green radial line in Figure \ref{fig:6}A corresponds to the intensity plot in Figure \ref{fig:6}B. The bell-shaped curve results from the radial line extending from within the cavity, passing through the high-density ring of the island, and ending at the noise floor region beyond the feature. The pixel location of maximum intensity is then recorded and plotted back onto the WISPR-I image. The green dots in Panel C display the maximum intensity points selected from each radial line plot, and the blue ellipse is the final, fitted ellipse. 

Occasionally, due to any bright object, say a star, passing behind the island, the wrong maximum intensity points may be selected. Therefore, a five-point smoothing algorithm is performed on the location of the maximum intensity points selected from the radial profile line plots. After selecting the point of maximum intensity, the best-fit ellipse is calculated using a direct least-squares fitting algorithm developed by \cite{1999Fitzgibbon} \cite[subsequently improved upon by][]{1998HalirandFlusser}. The fitting is approached by minimizing the general conic second-order equation in the standard form,

\begin{equation}\label{eq:general conic}
F(x_{px},y_{px}) = Ax_{px}^2 + Bx_{px}y_{px} + Cy_{px}^2 + Dx_{px} + Ey_{px} + G = 0,
\end{equation}
where $A$, $B$, $C$, $D$, $E$, and $G$ are the parameters and $x_{px}$ and $y_{px}$ are the pixels in the WISPR-I image. In order to ensure that the solution is an ellipse, the direct least-squares fitting algorithm imposes the restriction of $4AC - B^2 = 1$. For more details, refer to \cite{1998HalirandFlusser} and \cite{1999Fitzgibbon}. The algorithm used here is a Python version of the published Matlab code\footnote{Matlab code retrieved from \url{https://scipython.com/blog/direct-linear-least-squares-fitting-of-an-ellipse/.}} by \cite{1998HalirandFlusser}. With the parameters obtained through the direct least-squares fitting algorithm, we then calculate the major axis, minor axis, aspect ratio, rotation, and center point of the ellipse within the image. The rotation refers to the counter-clockwise rotation of the major axis of the ellipse from the projection of \textit{Parker's} orbit plane onto the image. The calculations provide an idea of the orientation of the ellipse relative to \textit{Parker's} orbit plane, as well as an estimation of the radial width and vertical thickness of the feature, after converting from pixel positions to HCI Cartesian coordinates.

There are two points of note regarding the Max-Point algorithm: First, the 1-degree interval between the radial lines will result in under-sampling the leading and trailing edges compared to the upper and lower edges. This occurs because the arc length covered by 1-degree increments of a circle is different than that of an ellipse, and the island has a more elliptical shape at the beginning of its propagation. Second, it is possible that the intensity value at the starting point within the selection box has the highest intensity value. This may occur if the origin point lands on a star or if the radial line happens to cross the island boundary in a region that has been artificially removed due to the LW-processing. In the latter case, no point for the radial line is chosen.

\subsection{Orientation of the Streamer via Gaussian Fit Algorithm\label{subsubsec:Gaussian Fit Algorithm}}
The Gaussian-Fit algorithm is designed to determine the rough angular position of the streamer center relative to \textit{Parker's} orbit plane. This algorithm calculates the best fit of a Gaussian to the intensity profile along a vertical line, aligned along the HP-longitude direction, that passes through the streamer. This algorithm is performed on WISPR-I L3-processed images with an applied sigma and uniform filter. 

\begin{figure}
    \centering
    \includegraphics[width=\textwidth]{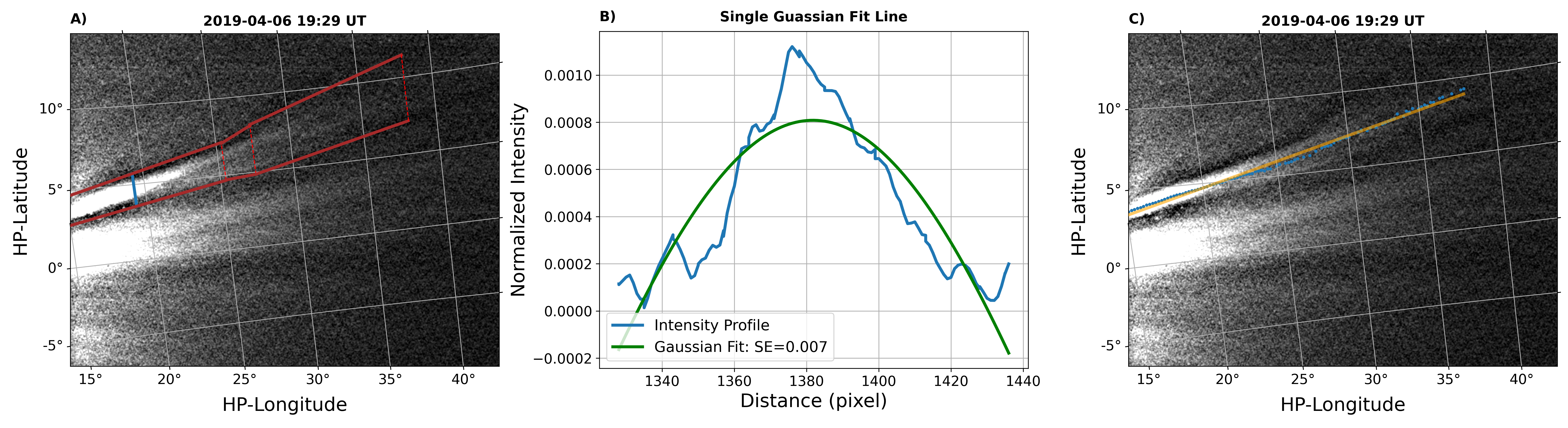}
    \caption{\textit{Panel A:} Sample vertical line (i.e., along a constant HP-longitude of $\sim19^{\circ}$) marked in blue within the bounded streamer lines (solid red). The different sections of the streamer are separated by dashed red lines. \textit{Panel B:} Resulting intensity profile along the blue vertical line (blue) with a Gaussian fit (green). The standard error of the Gaussian model and the actual data is $0.007$. \textit{Panel C:} The location of the intensity peak of each Gaussian curve (blue dots) plotted back onto the WISPR-I image, where the orange line is the linear fit of all the Gaussian peaks.}
    \label{fig:7}
\end{figure}

Due to the streamer pinch described in Section \ref{sec:WISPR Data Products and Initial Observation} and the presence of the magnetic island propagating along the streamer, the width of the streamer does not appear consistent through a single WISPR-I image. Therefore, we manually inspect each image and determine upper and lower bounds to isolate the streamer from the image background and create an area to sample the streamer. Figure \ref{fig:7} Panel A displays a sample of the upper and lower bounds, represented by solid red lines, and the vertical red dashed lines separating the sections of the bounded area on 2019 April 06 at 19:29 UT.

Once we establish the streamer bounds, we create vertical lines between the upper and lower bounds that sample the intensity of the streamer. The vertical lines are aligned along constant HP-longitude lines. Panel A in Figure \ref{fig:7} displays an example blue vertical line, located at $\sim19^{\circ}$ HP-longitude, that corresponds to the blue intensity profile in Panel B. The bell-shaped curve in Panel B results from the intensity along the vertical line that starts below the bright streamer, passes through the brightest portion of the streamer, and then ends above the streamer. The green line is the Gaussian fit to that intensity curve. The blue points in Panel C display the peak of each Gaussian fit plotted back onto the image. We use the location of the peak points to interpolate a line of best fit (orange), which is used to calculate the angle of the streamer relative to \textit{Parker's} orbit plane. We repeat this algorithm for each WISPR-I image from 18:14:25 UT on 2019 April 06 to 00:58:34 UT on 2019 April 07. 

The streamer becomes dimmer further from the Sun, which makes the vertical intensity profile flat. When this occurs, the Gaussian is no longer an ideal fit. Therefore, we create a statistical-based approach to determine when the Gaussian is no longer a proper fit to the intensity plot, for which we compare the standard error between the actual intensity and the predicted intensity of both a Gaussian fit and a straight-line fit along the vertical line sample. If the standard error of the Gaussian fit is within $10\%$ of the standard error of the straight-line fit, then the vertical line sample is discarded, and we move to the next vertical line sample. Alternatively, if the standard error of the Gaussian fit is less than that of the straight-line fit by more than $10\%$, then the selection of a Gaussian fit is considered valid, and the location of the maximum of the Gaussian is mapped back onto the WISPR-I image.

\section{Results and Discussion}\label{sec:Results and Discussion}

\subsection{Results in the WISPR FOV}\label{subsec:WISPR Results}

The magnetic island is observed from 2019 April 06 at $\sim$21:02:42 UT to 2019 April 07 at ~$\sim$00:58:34 UT in the WISPR-I images. The island feature appears to travel radially outward from the Sun, roughly perpendicular to WISPR-I's line of sight, as indicated by Figure \ref{fig:4}A. The center of the island feature travels approximately $7$ R$_{\odot}$, from $16.8$ to $23.5$ R$_{\odot}$ during the observation period, and the distance from the island center to \textit{Parker} decreases from $\sim$29 R$_{\odot}$ to 26 R$_{\odot}$. Using Equation \ref{eq:Viewing Angle}, we calculate that the viewing angle between \textit{Parker} and the island's center decreases by $\sim10^{\circ}$, from $113.06^{\circ}$ to $103.49^{\circ}$, during the island's observation period, supporting the propagation of the island as seen in Figure \ref{fig:4}A.

\begin{figure}
    \centering
    \includegraphics[width=0.8\textwidth]{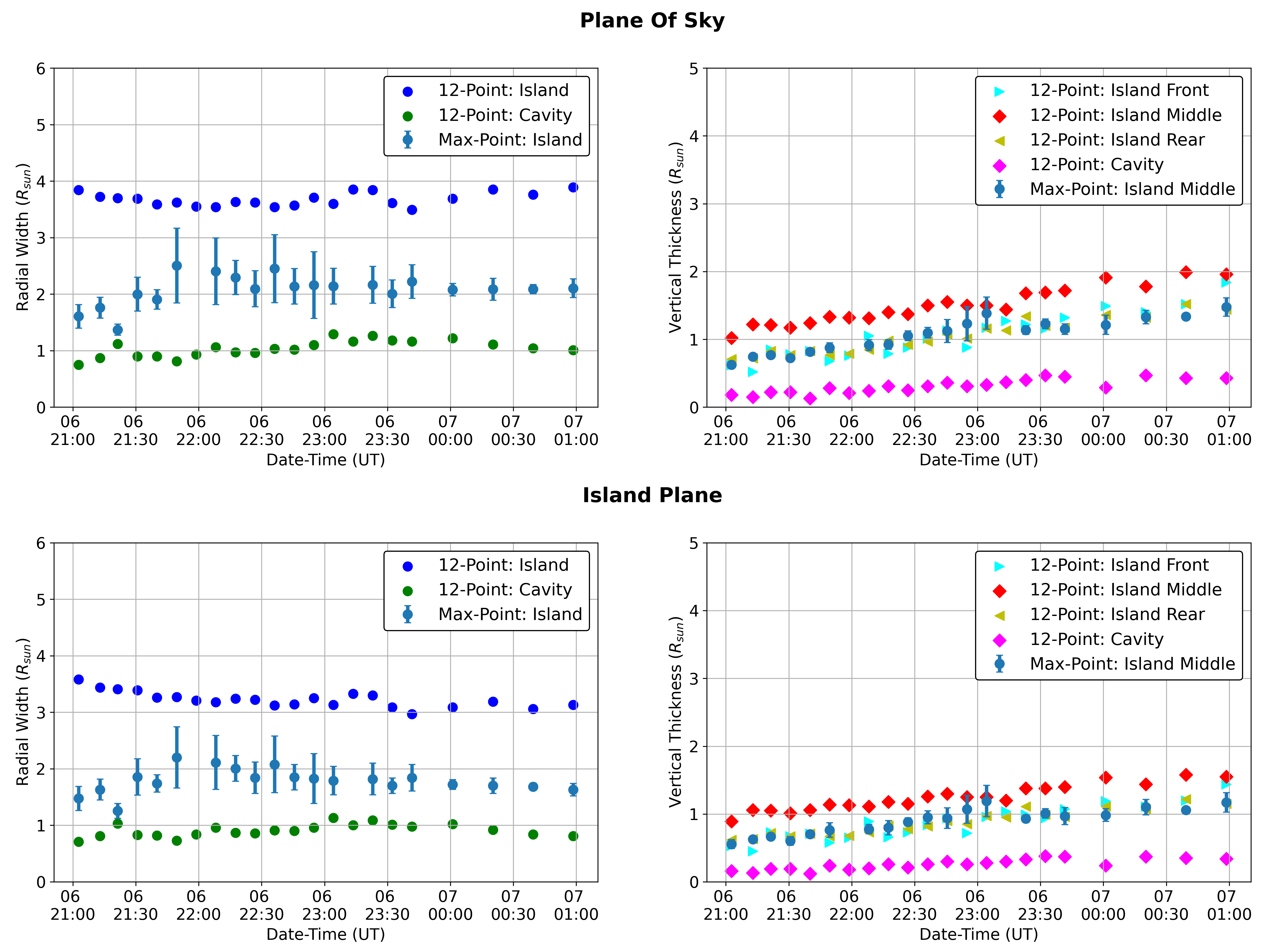}
    \caption{The dimensions of the island feature and cavity, as calculated by the 12-Point method and Max-Point algorithm in the plane-of-sky (top panels) and island plane (bottom panels). The left plots display radial width values, where the island (dark blue) and cavity (green) are provided by the 12-Point tracking method, and the island (blue with error bars) is provided by the Max-Point algorithm. The right plots display the vertical thickness values, where the island (yellow, red, cyan) and cavity (pink) are provided by the 12-Point tracking method, and the island (blue with error bars) is provided by the Max-Point algorithm. All distances are measured in R$_{\odot}$.}
    \label{fig:8}
\end{figure}

Figure \ref{fig:8} displays the evolution of the island's radial width and vertical thickness over the observation period. The plots display calculated dimensions from both the 12-Point method and Max-Point algorithm in the WISPR-I plane-of-sky (top two panels) and island plane (bottom two panels). We display the vertical thickness measurements of the island, as calculated from the 12-Point method, for the front, middle, and rear portions of the island and the center portion of the cavity (as denoted in the legend of the panel). The Max-Point method only measures the vertical thickness of the roughly center portion of the feature. The 12-Point and Max-Point results indicate that the radial width of the island and cavity experience minimal overall change (as seen in the left panels of Figure \ref{fig:8}), though the vertical thickness of the feature does experience a noticeable increase of $\sim 50-100\%$ R$_{\odot}$ by the end of the observation period (as seen in the right panels). Note that there is roughly a factor of two to three between the radial width values of the 12-Point and Max-Point results - this is because each method measures the island in a different way. The 12-Point method tracks the outer edges of the density ring (see Figure \ref{fig:3}), whereas the Max-Point algorithm measures the brightest points of that density ring (see Figure \ref{fig:6}). Though the 12-Point tracking method and the Max-Point ellipse algorithm do not measure the same sections of the island feature, both indicate that the magnetic island becomes more circular by the end of the observation period in both the plane-of-sky and island plane.

\begin{figure}
    \centering
    \includegraphics[width=0.6\textwidth]{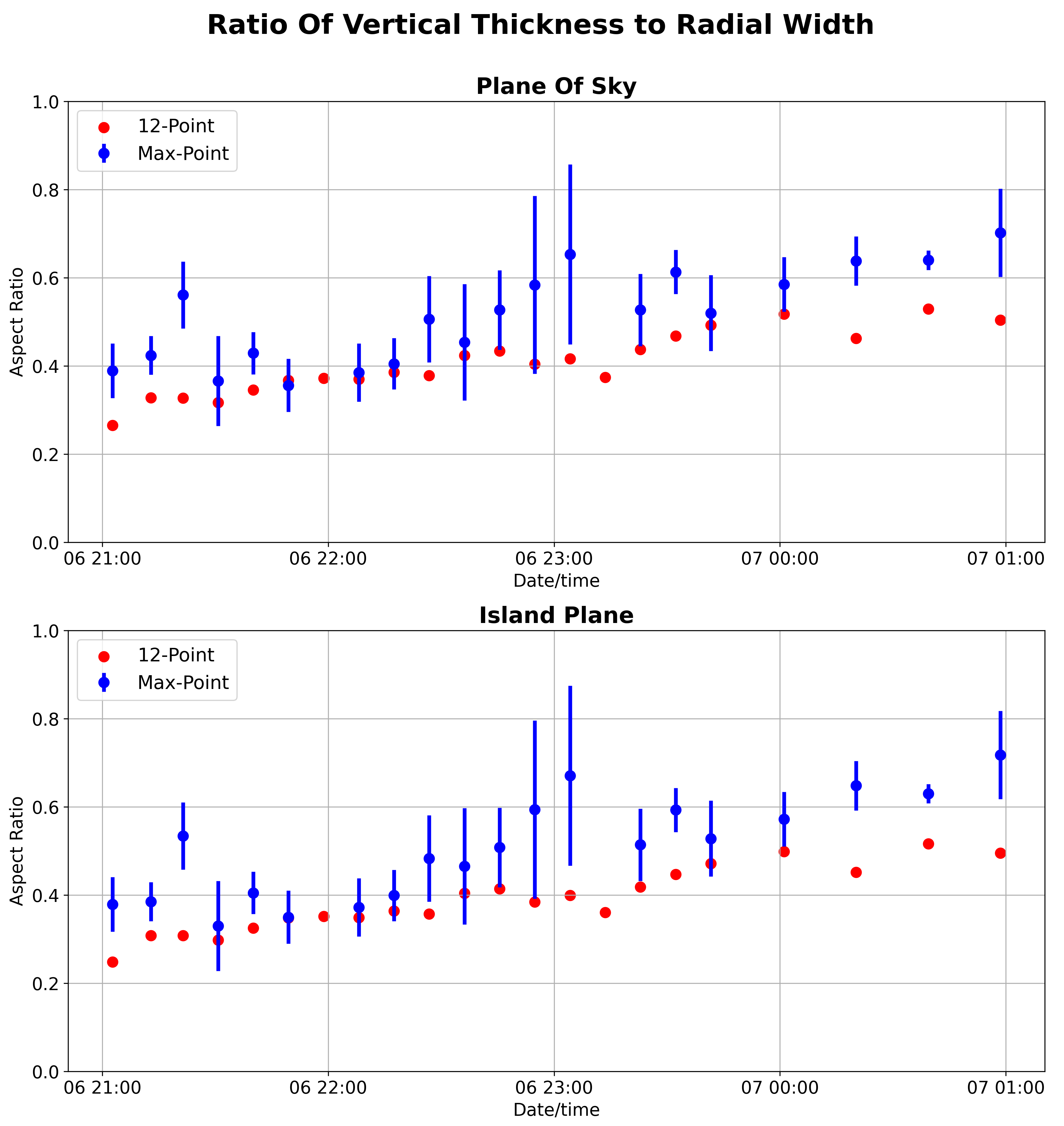}
    \caption{Comparison of aspect ratios, i.e., the ratio of the vertical thickness to the radial width of the island, as calculated from the 12-Point measurements (red) and the Max-Point measurements (blue). The top plot displays the aspect ratio measurements in the WISPR-I plane-of-sky, and the bottom plot displays the aspect ratio measurements in the island plane. The error bars of the Max-Point ellipse depict a range of two standard deviations.}
    \label{fig:9}
\end{figure}

The aspect ratio, calculated as the ratio of vertical thickness to radial width, increases towards 1 for both the 12-Point and Max-Point methods. In Figure \ref{fig:9}, we display those ratios for the plane-of-sky (top panel) and island plane (bottom panel) and find that, for both methods, the value of the aspect ratios approximately double over the observation period. This trend correlates to that observed in the WISPR-I images and the results presented in Figure \ref{fig:8}, where the radial width of the island appeared to experience minimal change over time while the vertical thickness increased, resulting in the island evolving from elliptical to circular shape.

Overall, both the island and cavity experience minimal change in their radial width and appear to have similar patterns of evolution. From these results, we infer that the cavity is physically within the outer density ring and is itself part of the whole island feature. If there were large differences in the evolution of their radial width or vertical thickness, then that may indicate that the cavity was rather a processing effect. However, as both the ring of enhanced density and the core of relatively lower density from the simulations in \cite{2005Rappazzo} match the form of the island discussed here, we propose that a plausible reason for this core density deficit may result from a buildup of magnetic guide field (field-aligned perpendicular to the 2.5-D plane of the simulation) at the center of the island. The simulations in \cite{2005Rappazzo} were initialized with a strong guide field concentrated at the current sheet, and so this field should be pulled into the center of the island early in the reconnection process while field lines initially near the current sheet/concentrated guide field region reconnect. The magnetic pressure of this concentrated guide field would then be able to support the center of the island against the magnetic tension that would build up around the island core during the subsequent phases of reconnection. Later sets of reconnecting field lines, which would come from farther away from the current sheet and so have little guide field, would not have this guide field pressure and would compress to a higher plasma density. Thus, we hypothesize that a visible cavity in an observed island may be a signature of a build-up of a guide field at the center of that island. In the \cite{2005Rappazzo} 2.5-D simulation, the presence of a guide field appears as reduced density in the center of the simulated islands. We infer that the evolution of the cavity, coupled with the inferred viewing angle between the island plane and the WISPR-I line of sight to the island center being close to $90^{\circ}$, indicate that we may, in fact, be looking along the axis of the island, i.e., along the axis of a 3-D flux tube formed by current sheet reconnection. This reinforces the presence of the cavity as a part of the whole magnetic feature versus just a processing effect.

We examine the kinematics of the magnetic island through the two tracking processes discussed in Section \ref{sec:Tracking Methods}. The T\&F method calculates a velocity of $\sim327 \pm 4.8 \unit{km\,s^{-1}}$ in the radial direction corresponding to the center of the island \cite[discussion of this calculation can be found in][]{2020Liewer}. Using the method discussed in Section \ref{subsubsec:Size and Kinematics Calculation}, we calculate 12 velocity and acceleration values that correspond to each tracked portion of the island feature. We average the 12 values to obtain an overall velocity and acceleration of $\sim340.43 \unit{km\,s^{-1}}$ and $-1.27 \unit{m\,s^{-2}}$ in the island plane, respectively. The velocity values of the two tracking methods give similar values, and the change in velocity calculated for the 12-Point tracking during the observation period is within the uncertainty of the velocity provided by the T\&F solution. The uncertainties between the two velocity values are small, so we believe that no significant uncertainties are introduced. Furthermore, an assumption of linear motion is acceptable here as the calculated acceleration is very small, and the time frame under consideration is relatively short. Both velocities are also similar to the velocity of streamer blobs and the outward component of in-out pairs, which have been observed over a range of $\sim$ $200-400 \unit{km\,s^{-1}}$ near $25$ R$_{\odot}$ \citep{1997Sheeley, 2002Sheeley&Wang, 2009Sheeley}. The velocity of the island feature is also similar to that of the slow solar wind, which has been found to be below $\sim$ $450 \unit{km\,s^{-1}}$ \citep{1990Schwenn}. However, the island's deceleration is modestly different than that of typical streamer blobs, which have accelerations between $3.4-5.5 \unit{m\,s^{-2}}$ \citep{1998Wang}. 

\begin{figure}
    \centering
    \includegraphics[width=.7\textwidth]{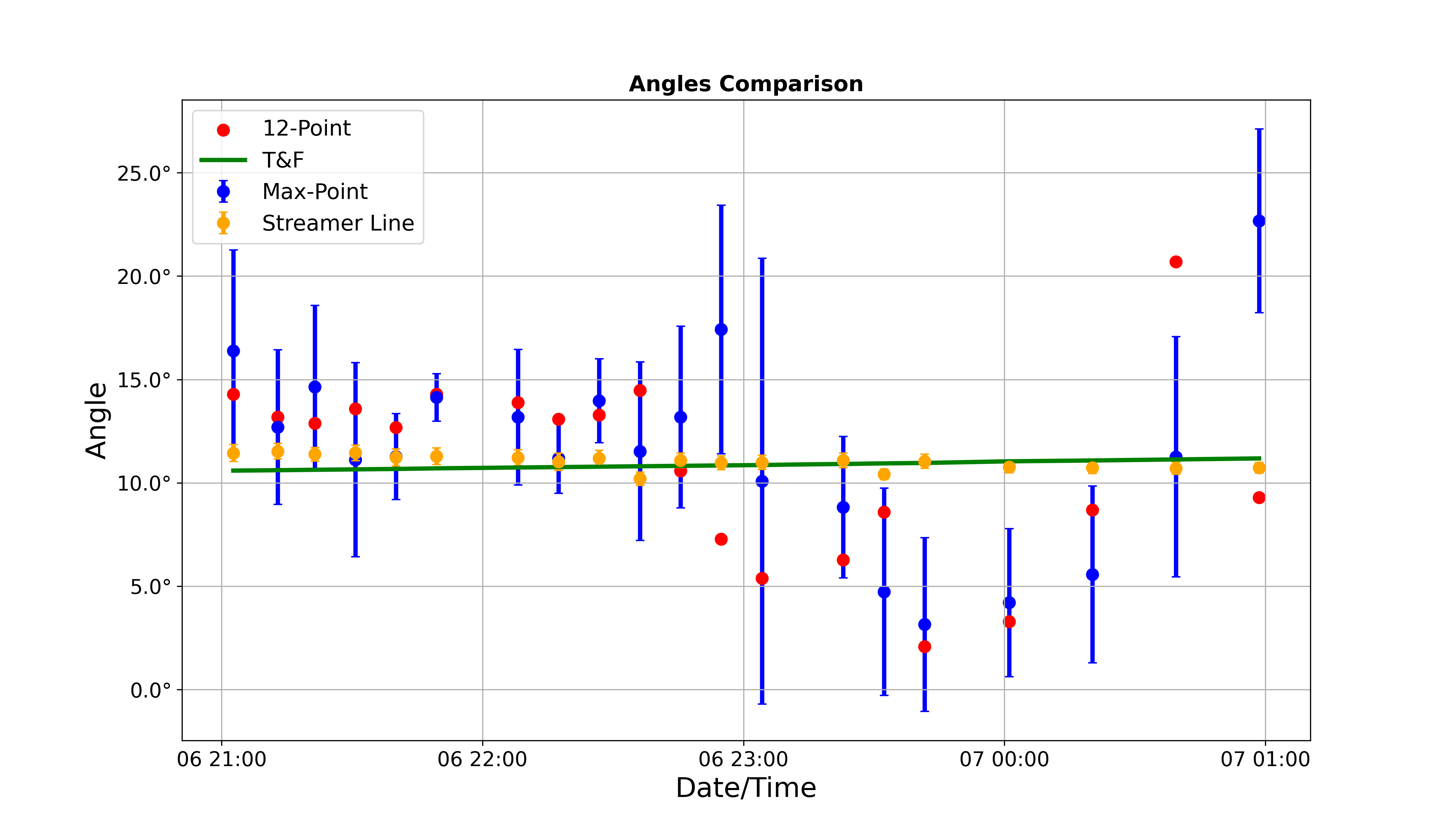}
    \caption{Orientation of the ellipse major axis provided by the 12-Point method (red) and Max-Point algorithm (blue) compared to the position of the T\&F ray (green) and streamer position (yellow). All angles are taken relative to \textit{Parker's} orbit plane.}
    \label{fig:10}
\end{figure}

Density enhancements such as streamer blobs have been found to trace the solar wind. So, as we cannot say from the velocity or acceleration values alone if this island traces the streamer, we inspect the island's orientation relative to the streamer using the Gaussian-Fit algorithm discussed in Section \ref{subsubsec:Gaussian Fit Algorithm}. We calculate the streamer's orientation relative to \textit{Parker's} orbit plane and compare this to the angle that the island's major axis makes relative to \textit{Parker's} orbit plane. We use the radial widths resulting from the 12-Point tracking and Max-Point algorithm to gauge if the island's major axis is oriented similar to the streamer. Figure \ref{fig:10} displays the 12-Point and Max-Point ellipse angles during the observation period compared to the streamer angle and T\&F solution. We find that the streamer maintains a steady inclination of $\sim11.04^{\circ} \pm 0.18^{\circ}$ relative to \textit{Parker's} orbit plane. The streamer angle is similar to the angle of the island propagation ray provided by the T\&F solution when projected on the WISPR-I images. The ray (red line in Figure \ref{fig:5}) makes an angle of $\sim10.83^{\circ}$ relative to \textit{Parker's} orbit plane. The 12-Point tracking results indicate that the island feature is oriented at an average angle of $\sim10.88^{\circ}$, and the Max-Point ellipse results indicate that the island feature is at an average angle of $\sim9.75^{\circ} \pm 2^{\circ}$. The 12-Point results are within the error bars of the Max-Point ellipses results, with a few exceptions towards the end of the observation period after 23:00 UT on 2019 April 06, likely due to the island becoming more circular, so what is understood as the major axis of the feature becomes more subjective. The island feature appears to have no effect on the streamer orientation as the streamer maintained a roughly consistent angle of $\sim11.04^{\circ}$ relative to \textit{Parker's} orbit plane with no drastic shift in position, nor did the streamer appear to split. The only notable effect on the streamer appears to have occurred before the island is visible in the WISPR-I images when the streamer pinch occurred at $\sim$ 16:12:52 on 2019 April 06, as shown in Figure \ref{fig:1}. 

\begin{figure}
    \centering
    \includegraphics[width=\textwidth]{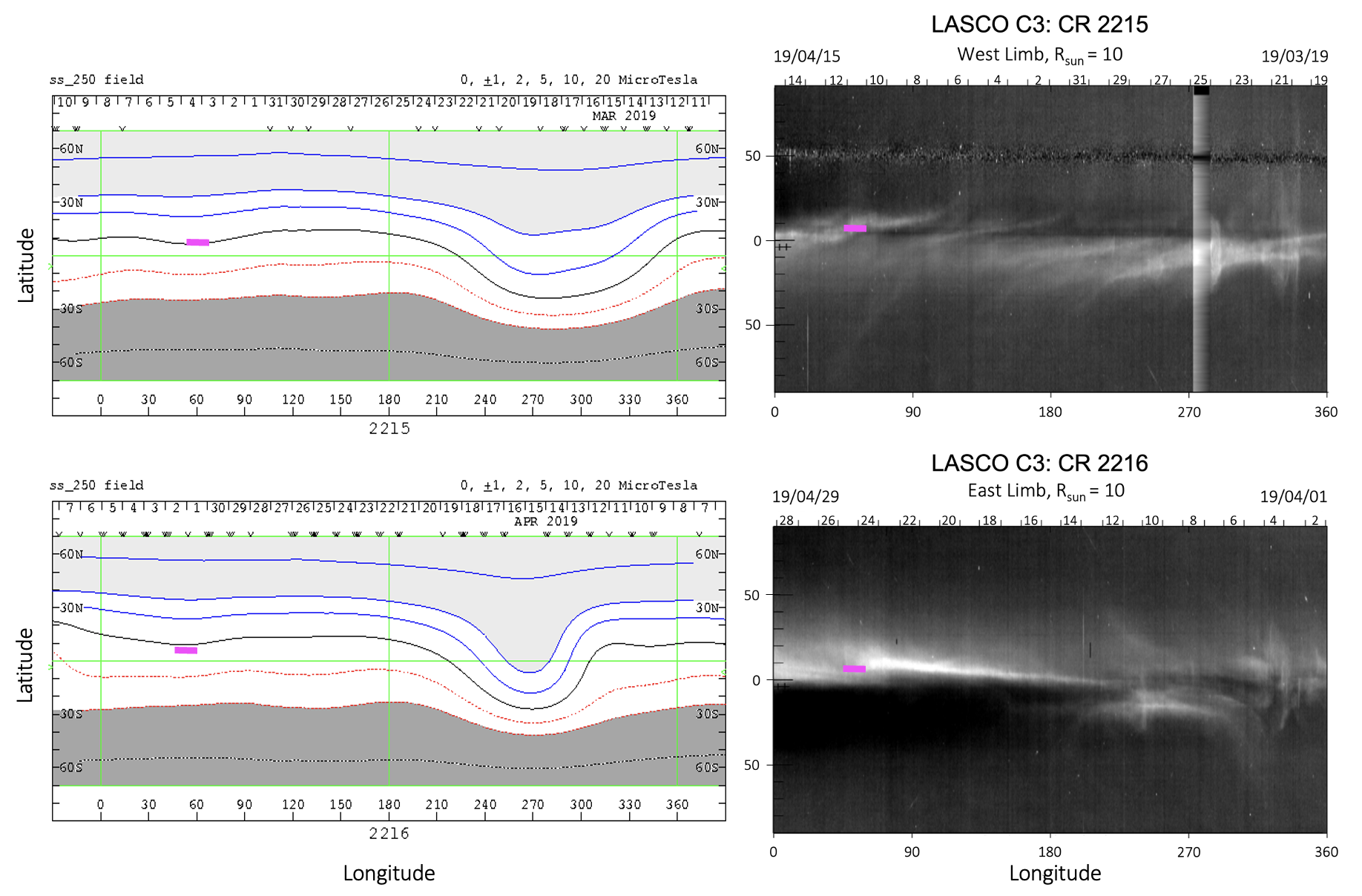}
    \caption{\textit{Left:} WSO synoptic charts with contours for the strength and sign of the radial magnetic field. The black contour represents zero radial field, i.e., the HCS. \textit{Right:} LASCO synoptic maps taken at $10$ R$_{\odot}$. The HCS appears as the elongated bright regions, and the magenta bar represents the magnetic island location at Carrington longitude and latitude = ($50^{\circ}$, $7^{\circ}$), which is close to the HCS (WSO map) and close to the helmet streamer (LASCO map). See also \cite{2024Liewer}, submitted, for a similar map for this island and three other streamer blobs.}
    \label{fig:11}
\end{figure}

The ellipse angle results relative to the streamer, coupled with the average velocity and acceleration of the island, indicate that the island is likely entrained in the slow solar wind. To investigate this further, we plot the island's position on the LASCO synoptic map and Wilcox Solar Observatory \cite[WSO;][]{1969Schatten, 1983Hoeksema} synoptic chart. Figure \ref{fig:11} displays the WSO\footnote{WSO source surface synoptic charts were retrieved from \url{http://wso.stanford.edu/synsourcel.html}.} (left) and LASCO\footnote{LASCO synoptic maps are a standard data product from the Naval Research Laboratory. The maps used here were retrieved from \url{https://lasco-www.nrl.navy.mil/carr_maps/c3/}.} (right) synoptic maps around the date of observation of the magnetic island feature, marked with the location of the magnetic island at Carrington longitude and latitude = ($50^{\circ}$, $7^{\circ}$) on each map. We found that the island location does fall near the heliospheric current sheet (HCS), which is represented by the black line in the WSO maps and is associated with the helmet streamer marking the HCS. The HCS is the boundary separating oppositely directed radial magnetic fields in the heliosphere \citep{1978Smith, 2001Smith}. The island falls near the tip of a very bright coronal helmet streamer, which is observed as the bright streak in the Carrington Rotation (CR) 2216 LASCO synoptic map. The two synoptic plots do not provide insight into the formation of the island, but the location of the island near the HCS and in association with a helmet streamer leads us to infer that the formation of the magnetic island occurs in the current sheet via magnetic reconnection, similar to streamer blobs and in-out pairs. We infer that the formation of this magnetic island occurs in the current sheet via magnetic reconnection at two locations between open field lines and that the island is a subset of streamer blobs.

\begin{figure}
    \centering
    \includegraphics[width=0.6\textwidth]{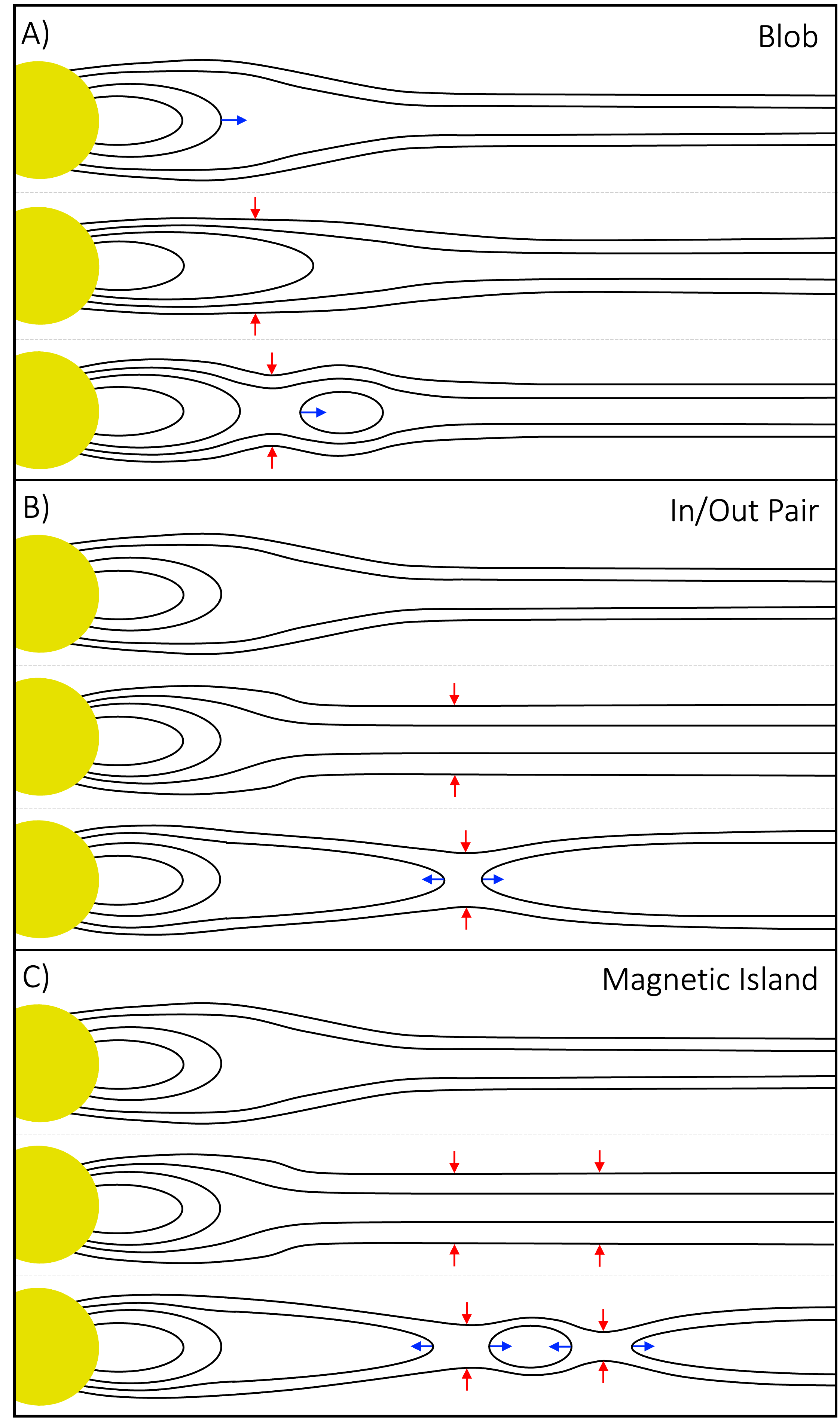}
    \caption{Diagram of the possible 2-D scenarios of the formation of streamer blobs, in-out pairs, and magnetic islands. Black lines represent magnetic field lines, the red arrows display reconnection sites, the blue arrows indicate the motion of the feature, and the Sun is represented by the yellow disk. Panel A is based on the streamer blob formation discussed in \cite{2018Higginson}, and Panel B is based on the in-out pair formation discussed in \cite{2009Sheeley, 2017Sanches-Diaz}. Panel C is based on the theory of magnetic island formation in a current sheet as discussed in \cite{2005Rappazzo}.}
    \label{fig:12}
\end{figure}

Figure \ref{fig:12} presents a diagram of the processes thought to be acting to form streamer blobs, in-out pairs, and the magnetic island. Panel A displays the formation of streamer blobs at the tip of a helmet streamer, as discussed in \cite{1997Sheeley, 2018Higginson}, where magnetic reconnection is caused by instabilities within the current sheet. This leads to a pinch (red arrows) that forms a blob at the tip of the helmet streamer, which then accelerates away from the Sun. Panel B displays the formation of in-out pairs, which are thought to form from reconnection well above the helmet streamer \citep{2009Sheeley, 2017Sanches-Diaz}. The sheet pinches (red arrows), causing an oppositely directed magnetic field on either side of the current sheet to reconnect, generating outflow jets to the left and right of the reconnection region. This results in two density enhancements at the outward and inward edge of a depleted region (the reconnection site). The edges separate (indicated by the blue arrows), and the outward component moves radially away from the Sun, while the inward component appears as a collapsing magnetic loop.

Panel C displays the formation of the magnetic island between two reconnection sites (indicated by the red arrows) well above the helmet streamer. The sheet pinches at two points, for example via a tearing mode instability \citep{1963Furth}, leading to reconnection at those two points. This generates two pairs of outflow jets on either side of each reconnection site. This forms an inward-flowing loop below the inner pinch-point, an outward-flowing loop above the outer pinch-point, and an island between the two pinch-points. The inward and outward-flowing loops are analogous to the in-out pairs in Panel B. While we do not see evidence of an inward-flowing loop (this may be below the field of view of WISPR-I), we do see the outward-flowing loop (indicated by the cyan arrow in Figure \ref{fig:1}) and the island (indicated by the green arrow in Figure \ref{fig:1}). In a symmetric reconnection event with no external flows, the island would be stationary. But, in the case we focus on here, the island was presumably formed in the current sheet and dragged by the out-flowing solar wind and so continues to move with the wind (not shown in the figure).

\subsection{Results in Other FOVs}\label{subsec:Other Results}

COR2A and C2 coronagraphs do not observe any significant presence of the magnetic island on the respective days of observation. We build on the exercise from Section \ref{subsec:Tracking and Fitting Method} and Figure \ref{fig:2} by taking a closer look at the island propagation during the period for which the island's path is within the COR2A FOV, which we determined to be 18:24:00 – 22:39:00 UT on 2019 April 06. Within this period, we calculate that the approximate distance from STEREO-A (COR2A) to the inferred island center is $219$ R$_{\odot}$ and the viewing angle between the island plane and line of sight from COR2A to the island decreases slightly, from $52.78^{\circ}$ to $51.45^{\circ}$. This slight change in viewing angle supports what we see in Figure \ref{fig:4}A, in that the island moves away from STEREO and is not perpendicular to STEREO's line of sight. In the COR2A images, we do not see a distinct island feature as shown in Figure~\ref{fig:2}; however, we do notice a slight change in the shape of the streamer coinciding with the location of the island coordinates. As we cannot distinguish a feature within these images, we can only infer that the island feature may be moving within this streamer, which could have caused the change of shape. As noted in Section \ref{sec:WISPR Data Products and Initial Observation}, the streamer changes shape before the island is visible and once the island feature travels along the streamer (see Figure \ref{fig:1}). This slight bulge of the streamer is visible in the COR2A images at the location of the island, though no streamer pinch is visible in the COR2A images as it is in the WISPR-I images.

We determined the time range for which the island's path is within C2's field of view to be 16:00:05 - 18:24:05 UT on 2019 April 06, during which the viewing angle between the island plane and line of sight from C2 to the island decreases slightly, from $148.38^{\circ}$ to $147.88^{\circ}$. The island is approximately $199$ R$_{\odot}$ from LASCO and appears to move towards the satellite, as seen in Figure \ref{fig:4}A, at a much larger viewing angle than that between the island and \textit{Parker}. We plotted the calculated coordinates for the time frame 16:00:05 – 18:24:05 UT on 2019 April 06 and saw no discernible feature present at the location of the island coordinates. This is to be expected as the island is moving towards the spacecraft, so, though we do not know what the feature looks like in the third direction, it would likely not resemble that of an island-like or blob-like feature.

\begin{figure}
    \centering
    \includegraphics[width=\textwidth]{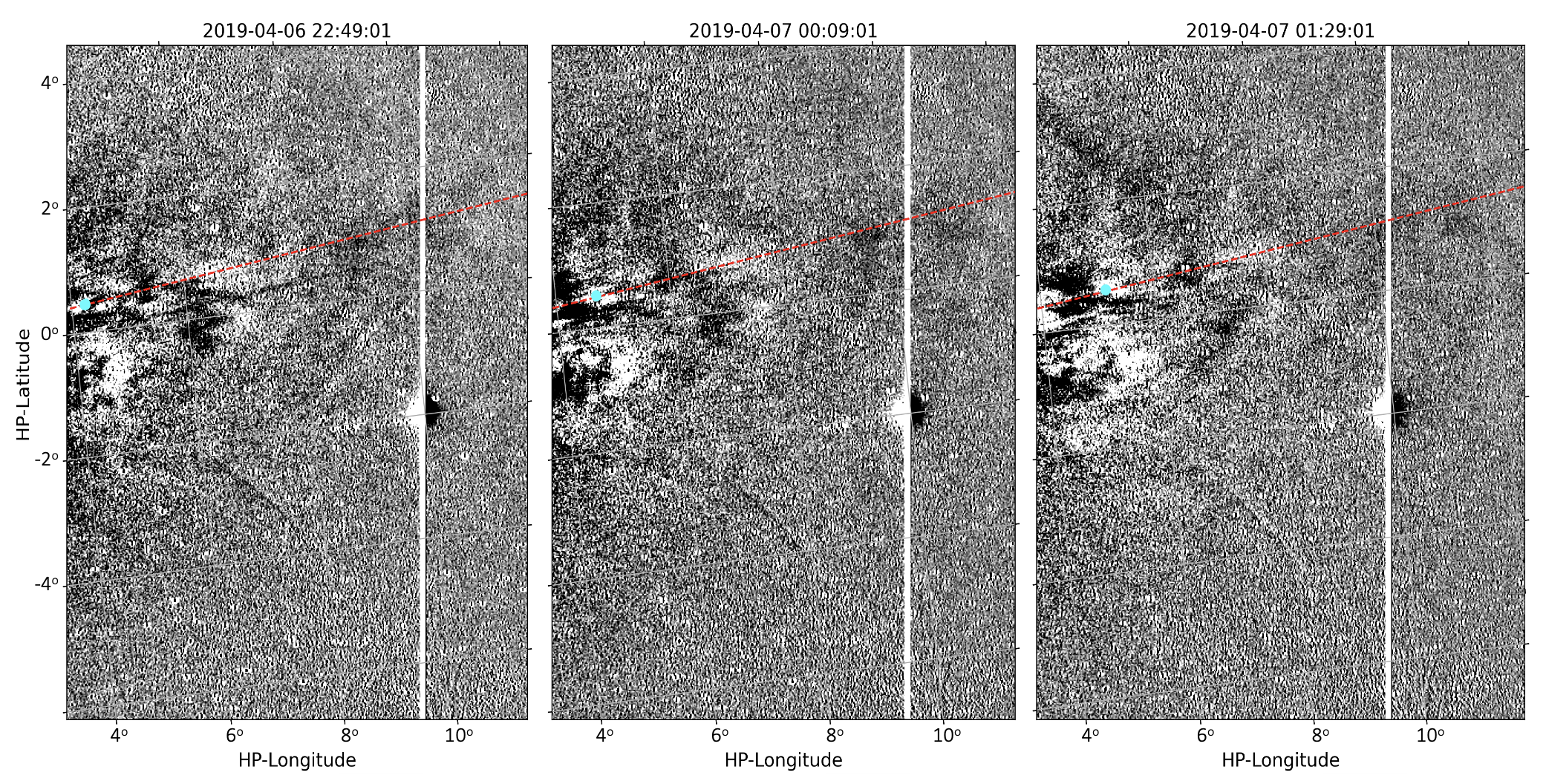}
    \caption{Calculated coordinate overlay on HI1A difference images. The cyan dot is the projected location of the calculated 3-D island coordinate, and the red dashed line is the HCI longitude and latitude, provided by the T\&F trajectory method, projected onto the HP-longitude/latitude coordinate plane. The cyan dot appears to follow a specific density enhancement, though there is no visible cavity. The density enhancement becomes indistinguishable from other solar wind features by the end of the observation period displayed here.}
    \label{fig:13}
\end{figure}

HI1A, a much more sensitive camera than both C2 and COR2A, does show a density enhancement at the location of the island coordinates. Performing the same analysis process as COR2A and C2, we determine the time range for which the island's path is within HI1A's field of view to be from 22:09:01 UT on 2019 April 06 – 02:09:01 UT on 2019 April 07. Within this period, the distance from STEREO-A (HI1A) to the inferred island center is $\sim222$ R$_{\odot}$, and the angle between the island plane and the line of sight from HI1A to the island decreases slightly, from $51.63^{\circ}$ to $50.42^{\circ}$. The cyan dots in Figure \ref{fig:13} are plotted at the location of the 3-D calculated coordinates in the HI1A FOV. Many density enhancements are visible in the sequence of images, though the coordinates appear to roughly track one particular density enhancement. Upon further inspection, the density enhancement observed in HI1A appears to be less coherent of a structure when compared to the magnetic island observed by WISPR-I. Visually, the density enhancement does not appear to have a consistent elliptical shape or become circular during the observation period. Additionally, no visible cavity is present at any point during the feature’s propagation in the HI1A viewing period. It is important to note that HI1A is viewing the island from a different direction than WISPR and is likely not viewing the axis of formation of the island feature. If, in 3-D, the island is a flux rope, we could expect a dramatic change in its projected shape as the line-of-sight changes. This would result in the disappearance of a cavity if the line of sight is not closely aligned with the axis of the island/flux rope, as would be the case for this HI1A observation. Therefore, we would not expect to observe the cavity within the density enhancement but rather an oblique projection of the island, including whatever its structure along the line of sight of WISPR might have been. This argument holds for both C2 and COR2A as well.

We apply the 12-Point tracking method to track the density enhancement throughout the determined HI1A viewing time range. We track only four points for this feature: the distance between the feature’s leading/trailing (radial width) and upper/lower (vertical thickness) edges from time 22:49:01 UT on 2019 April 06 to 01:29:01 UT on 2019 April 07, as the entire feature is not in frame at the start of the observation period and is indistinguishable from other density enhancements by the end of the observation period. Using the methods discussed in \ref{subsubsec:Size and Kinematics Calculation}, we find that, in the island plane, the 12-Point tracking method (in fact, 4-points here) yields an average velocity of $270.76\unit{km\,s^{-1}}$ and acceleration of $4.38 \unit{m\,s^{-2}}$. We calculate that, in the HI1A plane-of-sky, which is at $\sim$100 degrees HCI longitude, the radial width of the density feature increases from $1.11$ to $1.28$ R$_{\odot}$ and the vertical thickness increases from $0.76$ to $0.83$ R$_{\odot}$ by the end of the HI1A observation period. In the island plane, the radial width of the density feature increases from $1.15$ to $1.33$, and the vertical thickness increases from $0.79$ to $0.87$, respectively. The dimensions of the density feature tracked in the HI1A images does not evolve in a similar fashion to the island observed by WISPR-I, though the kinematics of both the density feature and island are similar to that of streamer blobs. The difference in dimension may, again, be due to the fact that HI1A is viewing the island from a different direction than WISPR, so we would only see a projection of the island feature. We also cannot rule out the possibility that motion blurring in the HI-1 images has erased any signatures of a faint cavity, as the HI-1 images are an on-board sequence of individual exposures obtained over a 20-minute period. For these two reasons, we would not expect to see a cavity within this density feature.

Several factors likely play into the lack of observation of the magnetic island by COR2A and C2, as well as the lack of a visible cavity in HI1A, but foremost among those is simply the visual brightness (or lack thereof) of the island. To demonstrate this, we took the slightly unconventional approach of evaluating the apparent visual magnitude ($V$) of the brightest portions of the island as seen by WISPR-I, then comparing those values to the limiting magnitude for the C2 and C3 coronagraphs, as well as COR2A coronagraph and HI1A imager. While obtaining $V$ estimates of diffuse sources in WISPR observations is extremely challenging, an aperture photometry-based approach, similar to that used by \cite{2020Battams}, determined portions of the magnetic island to be approximately $V = 11$, albeit with substantial uncertainties. This is well above the limiting magnitude of $V\sim8.5$ for stellar sources in LASCO C3, and $V\sim9.0$ for LASCO C2, as well as $V\sim7.0$ for sources in STEREO COR2A, \textit{but} below the limiting magnitude of $V\sim13$ for sources in STEREO HI1A \citep{2017Battams&Knight}. 

As a rough validation of this approach, we performed an assessment of the brightness of a small CME observed in both LASCO and WISPR on 01 November 2018, when the instruments observing lines of sight were viewing roughly similar portions of the sky. Concurrent observations of this CME in both LASCO C3 and WISPR-I from 03:18 UT on 01 November 2018 to 21:45 UT on 02 November 2018 returned largely consistent visual magnitude estimates between the two cameras, up until the point the CME was no longer visually detectable in C3 ($V < \sim8.5$) even though it remained visible in WISPR-I and then WISPR-O for over 24 hours longer. Despite the large uncertainties in applying such photometry techniques to diffuse solar structures, these results returned a numerical approximation for what can clearly be observed by the eye - namely, the WISPR cameras are far more sensitive to faint coronal structures than the LASCO cameras. 

Thus, we are confident that the faint magnetic island feature observed on 2019 April 06 by WISPR-I is likely too faint to have been detected by C2 or COR2A and that HI1A's lower spatial resolution when compared to WISPR \citep{2016Vourlidas} is a factor as to why the magnetic island feature is not discernible in the heliospheric images. This result appears to indicate that both high photometric sensitivity and high spatial resolution, as well as coincidence between the line of sight and island axis, are important factors in observing a magnetic island such as this one.

\section{Summary and Conclusions}\label{sec:Summary and Conclusion}
In this paper, we have presented an observational study of a possible magnetic island feature captured by the WISPR-I instrument during the second encounter of the \textit{Parker Solar Probe} mission. We inspected the island feature in both L3- and LW-processed images during the period when the island was visible in WISPR-I images, from 2019 April 06 at 21:02:42 UT to 2019 April 07 at 00:58:34 UT, and noted a visible density deficit in the island's center. We utilized the T\&F solution, which provides the location and speed of the island feature, in our 12-Point tracking method, from which we calculate the speed, orientation, and geometry of the tracked island feature. We conducted an analysis to calculate the feature's orientation in 2-D space and its location in 3-D space, as well as an approximate physical size of the island feature in both the plane-of-sky and the island plane. We developed the Max-Point algorithm as a check against the 12-Point tracking size results and conducted a Gaussian-fit technique to estimate the orientation of the streamer in \textit{Parker's} orbit plane. The key results of this study can be summarized as follows:

\begin{enumerate} 
    \item The T\&F method provides the radial velocity, radial distance from the Sun center at the start of tracking, and the longitudinal and latitudinal position of the magnetic island center. From these solutions, we found that the island traveled approximately $7$ R$_{\odot}$, from $16.86$ to $23.51$ R$_{\odot}$, from 21:02:42 UT on 2019 April 06 to 00:58:34 UT on 2019 April 07 along HCI longitude of $148^{\circ}$ and HCI latitude of $7^{\circ}$. We calculated the angle between WISPR-I's line of sight to the island and the island's direction of propagation to be roughly perpendicular, decreasing from $\sim113^{\circ}$ to $\sim103^{\circ}$ during the observation period.
    \item The 12-Point method found that, in the WISPR-I plane-of-sky, the radial width of the island increased slightly from $3.84$ to $3.89$ R$_{\odot}$, and the vertical thickness increased from $1.02$ to $1.96$ R$_{\odot}$. The radial width of the cavity increased from $0.75$ to $1.01$ R$_{\odot}$, and the vertical thickness increased from $0.18$ to $0.43$ R$_{\odot}$. In the island plane, the radial width of the island decreased from $3.58$ to $3.13$ R$_{\odot}$, and the vertical thickness increased from $0.89$ to $1.55$ R$_{\odot}$. The radial width of the cavity increased from $0.71$ to $0.81$ R$_{\odot}$, and the vertical thickness increased from $0.16$ to $0.34$ R$_{\odot}$ by the end of the observation period.
    \item The Max-Point algorithm found that, in the WISPR-I plane-of-sky, the radial width of the island slightly increased from $1.61$ to $2.10$ R$_{\odot}$, and the vertical thickness increased from $0.62$ to $1.47$ R$_{\odot}$. In the island plane, the radial width of the island increased from $1.48$ to $1.63$ R$_{\odot}$, and the vertical thickness increased from $0.56$ to $1.17$ R$_{\odot}$ by the end of the observation period. The Max-Point algorithm calculates these distances based on a different part of the island than the 12-Point method, so the difference in the two results is expected.
    \item The calculated aspect ratio, taken as the ratio of vertical thickness to radial width, of both the 12-Point and Max-Point solutions increase towards 1 in both the WISPR-I plane-of-sky and the island plane. This trend indicates that the island changed from elliptical to more circular in nature during the observation period.
    \item The T\&F method provides a constant velocity of $v = 327 \unit{km\,s^{-1}}$, and the 12-Point method provides a projected velocity and acceleration of $340.43 \unit{km\,s^{-1}}$ and $-1.27 \unit{m\,s^{-2}}$ in the island plane, respectively. Both velocity values are similar to that of the slow solar wind, as well as previously observed streamer blobs and the outward component of in-out pairs.
    \item The streamer maintains an angle of $\sim11.04^{\circ}$ relative to \textit{Parker's} orbit plane, similar to the ray provided by the T\&F method, which is at an angle of $\sim10.83^{\circ}$ relative to \textit{Parker's} orbit plane. The orientation of the island feature's major axis produced by the 12-Point method and the Max-point algorithm are in good agreement and appear to be consistent with the streamer angle up until the island takes a circular shape. 
    \item There is no visible island-like feature at the inferred island location in the C2 or COR2A images; however, the streamer observed in the COR2A images experienced a slight change of shape where the island coordinates were located.
    \item There is a density enhancement at the location of the inferred island coordinates in the HI1A images, but there is no visible inner cavity or other distinguishing island-like features.
    \item We calculated the apparent brightness magnitude of the island feature in the WISPR-I image to be $V\sim11$, which is well above the limiting magnitude for sources to be detected by C2 and COR2A but below the limiting magnitude for sources to be detected by HI1A.
\end{enumerate}

Our results show similarities between the magnetic island presented in this study and previously observed streamer blobs, though we argue that this island feature may be a subset of the general streamer blob, as proximity and camera sensitivity appear to be key factors to observe this feature. Additionally, the orientation of the feature relative to the observing instrument’s line of sight would affect its projected shape, resulting in the partial or entire disappearance of the inner cavity. These results imply that HI1A may have previously observed similar magnetic island-like features, but the features were understood as general streamer blobs or density enhancements. A future study of discrete, small features observed by WISPR and compared to what is seen by HI1A FOV when both lines of sight are roughly aligned would provide insight into the various faint features being observed by the WISPR cameras.

Additionally, the Solar Orbiter Heliospheric Imager (SoloHI) onboard \textit{Solar Orbiter} has better sensitivity and spatial resolution than HI1A and will reach a minimum perihelion of $0.28$ AU, indicating that islands such as this one seen by WISPR-I should be detectable in SoloHI, as WISPR observed this island when it was at a distance of $\sim$ $0.18$ AU from Sun center. It would be well worth comparing observations of the same FOV made by both WISPR and SoloHI to provide observers insight into transient structures like the magnetic island feature in the inner heliosphere and further link remote sensing and in-situ observations.

\section*{Acknowledgments}
\textit{Parker Solar Probe} was designed, built, and is now operated by the Johns Hopkins Applied Physics Laboratory as part of NASA’s Living with a Star (LWS) program (contract NNN06AA01C). Support from the LWS management and technical team has played a critical role in the success of the \textit{Parker Solar Probe} mission. This work was supported by the NASA \textit{Parker Solar Probe} Program Office for the WISPR program (Contract NNG11EK11I) to NRL. The Wide-Field Imager for \textit{Parker Solar Probe} (WISPR) instrument was designed, built, and is now operated by the US Naval Research Laboratory in collaboration with Johns Hopkins University/Applied Physics Laboratory, California Institute of Technology/Jet Propulsion Laboratory, University of Gottingen, Germany, Centre Spatiale de Liege, Belgium and University of Toulouse/Research Institute in Astrophysics and Planetology. M.L.A., A.G., and S.B.S. would like to acknowledge the support provided by George Mason University via the NRL contract N00173-23-2-C603. The work of P.C.L. was conducted at the Jet Propulsion Laboratory, California Institute of Technology, under a contract from NASA. The data used in this study are publicly available via the WISPR Data Access on the Wide-Field Imager for \textit{Parker Solar Probe} Project Website, which can be found at \href{https://wispr.nrl.navy.mil/}{https://wispr.nrl.navy.mil/}.

\bibliography{Main}
\bibliographystyle{aasjournal}

\end{document}